# Full-Domain Coupler: A Wireless Native Neural Backbone for Channel Representation and Deduction

Zirui Chen [a,b†], Ziqing Xing [a,b,c†], Zhaoyang Zhang [a,b,c*], Hongning Ruan [a,b], Yuzhi Yang [d], Zhaohui Yang [a,b], Chongwen Huang [a,b], Mérouane Debbah [d]

[a] *College of Information Science and Electronic Engineering, Zhejiang University, Hangzhou 310027, China*
[b] *Zhejiang Provincial Laboratory of Multi-Modal Communication Networks and Intelligent Information Processing, Hangzhou 310027, China*
[c] *Institute of Fundamental and Transdisciplinary Research, Zhejiang University, Hangzhou 310058, China*
[d] *Khalifa University of Science and Technology, PO Box 127788, Abu Dhabi, UAE*

† Co-first authors.
* Corresponding author.
*E-mail address:* ning_ming@zju.edu.cn (Z. Zhang).
*Part of this work is to be presented at 2026 IEEE International Symposium on Personal, Indoor and Mobile Radio Communications (PIMRC)* [1]

## ARTICLE INFO

*Article history:*
Received xxx
Revised xxx

**ABSTRACT**
Data representation is a fundamental issue in deep learning. However, as wireless data scale and deeply couple across many physical domains such as time, space, and frequency, existing wireless artificial intelligence (AI) technologies lack dedicated representation solutions. Instead, they mainly rely on stitching general-purpose networks, a “tool-driven” paradigm that inevitably results in structural redundancy and bottlenecks in information flow. To fill this gap, this paper proposes Coupler, a wireless native-AI neural backbone designed for the representation learning of channel state information (CSI)—the pivotal data in wireless systems. Leveraging the physical insights revealed by channel tensors, Coupler decomposes representation learning into individual domains on a layer-by-layer basis and then couples the learned domain-specific features through a dimension-staggered cascade. This full-domain interleaved learning architecture enables superior parameter efficiency and fine-grained multi-domain feature fusion. Based on this backbone, we use complex-domain multilayer perceptrons (CMLPs) as spatial and frequency domain learners, while employing three optional mechanisms—convolution, attention, or gating—to capture temporal dependencies. This results in a series of efficient and extremely lightweight channel learning schemes with diverse functionalities, showcasing the compactness, versatility, and flexibility of Coupler. We evaluate these schemes on channel deduction, a general representation task encompassing channel estimation, interpolation, prediction, and feedback. Extensive experimental evaluations validate the significant performance gains and robust applicability, even for real-world measured data, demonstrating Coupler’s potential as a promising basic architecture in the design of wireless foundation models.

## 1. Introduction

The deep integration of artificial intelligence (AI) and communications has emerged as a defining trend in wireless evolution [2], [3], [4], [5], driving the transition toward AI-native sixth-generation (6G) networks [6], [7], [8]. Leveraging AI’s strong capacities, such as implicit feature extraction, high-dimensional representation, and adaptive decision-making, 6G is expected to support a myriad of AI-native applications with superior performance. Extensive studies have already validated the potential of AI in complex network scenarios, including intelligent resource scheduling [9], adaptive beam management [10], and autonomous network orchestration [11]. However, growing evidence suggests that directly porting general-purpose learning models from computer vision or natural language processing to wireless systems often yields suboptimal results in terms of computational efficiency, generalization, and reliability. These challenges motivate another inquiry within AI-native 6G: To fully unlock the potential of AI in wireless contexts, it is essential to account for the intrinsic characteristics of wireless systems and to develop native AI methods tailored to the *physical properties* of wireless data, the function orientation of wireless tasks, and the hardware constraints of wireless devices [12].

In wireless systems, channel state information (CSI) serves as a pivotal information carrier. It characterizes the propagation structure of the wireless environment shaped by path loss, scattering, and fading along the transmission link, providing a critical reference for state awareness and transmission configurations [13]. As wireless systems evolve toward 6G, larger-scale antenna arrays, wider communication bandwidths, higher user mobility, and emerging functional requirements such as integrated sensing and communication will further increase the dimensionality, dynamics, and functional richness of CSI. Consequently, conventional pilot-intensive channel acquisition techniques encounter severe scalability bottlenecks, making it essential to extract and utilize the intrinsic features of CSI more efficiently. Governed by fundamental physical laws, the degrees of freedom of high-dimensional CSI tensors are essentially determined by a finite set of spatial-geometric and electromagnetic parameters, meaning that they are intrinsically confined to a low-dimensional manifold residing in a high-dimensional space. Leveraging this property provides a strong theoretical and technical basis for efficient CSI representation learning and has driven extensive deep learning (DL)-empowered studies on CSI-centric tasks, mapping, feedback, and prediction [14], [15], [16], [17], as well as downstream applications such as user localization and environmental sensing [18], [19], [20].

Refs. [21], [22], [23] employed methods such as convolutional neural networks (CNNs) or long short-term memory (LSTM) to learn from time-frequency signals, achieving intelligent channel estimation. Furthermore, to address the spatial dimension expansion brought by massive multiple-input multiple-output (MIMO), Refs. [24], [25], [26] utilized multilayer perceptrons (MLPs), complex-domain MLP-Mixer, and neural ordinary differential equations (ODEs), respectively, to achieve channel mapping in the space-frequency domain, thereby significantly reducing the signal overhead in high-dimensional channel acquisition. Parallel to these advances, substantial research has focused on mitigating feedback overhead in frequency division duplex (FDD) mode through learning-based compression. Specifically, Ref. [27] proposed CsiNet, a CNN-based autoencoder architecture to compress CSI in the angular-delay domain. Building upon this foundation, Refs. [28], [29], [30], [31] further optimized multiple aspects of feedback networks, including compression efficiency, computational complexity, and feedback flexibility. To bypass real-time signaling constraints altogether, AI-driven high-precision channel prediction has also emerged as a promising frontier. This includes time-series forecasting models based on LSTM [32], [33], Transformer [34], or ODE-recurrent neural networks (RNNs) [35], as well as innovative position-to-channel prediction schemes that employ generative networks with sinusoidal activation functions [36].

Beyond its traditional role in link adaptation, CSI is increasingly recognized for its profound spatial-related attributes, which pave the way for integrated sensing and communication (ISAC). This has catalyzed extensive research into channel-to-spatial-state representation learning. Refs. [37], [38], [39], [40] explored how to increase the accuracy and flexibility of channel positioning by leveraging learning architectures such as CNNs, MLPs, LSTM, and Transformer. Furthermore, Ref. [20] demonstrated the diverse functional characteristics of CSI-based scenario-sensing schemes under varying network designs.

However, despite demonstrating the immense potential of DL, current implementations predominantly rely on repurposing or heuristically stitching popular general-purpose architectures for channel representation. Although these DL-based studies show preliminary data-driven gains, insufficient alignment with the high-dimensional tensor structure and multi-domain coupling of CSI limit efficiency and generalization, motivating wireless-native innovations such as task-driven [8], [41] and physics-inspired [16] learning methods. In addition, from the perspective of framework evolution, a defining trend is to maximize the extraction of structural features and intrinsic correlations. As a milestone in this direction, Ref. [42] presents the channel deduction framework. It employs an end-to-end AI model to represent the full current channel by fusing past channel samples with current pilot-based coarse estimates. This framework allows the learning process to be conducted collaboratively across as many fundamental dimensions as possible (e.g., time, space, and frequency), providing a broader information interaction space. However, a fundamental gap remains: Architectures migrated from non-communication fields are not inherently designed to handle the growing physical dimensions. Consequently, even advanced models such as vanilla channel deduction neural networks (CDNets) [42], [43] often resort to the piecemeal integration of backbones such as LSTM, Transformer, and Mixer to handle temporal and spatial-frequency features separately. This "tool-driven" paradigm inevitably introduces computational redundancy and restricts the efficacy of inter-domain collaborative learning. In other words, there is an imperative need for a wireless-native backbone network—one that is fundamentally guided by wireless characteristics and capable of seamlessly integrating a wide array of physical dimensions.

To bridge this technical gap, this paper proposes Coupler, a neural backbone based on full-domain interleaved learning (FDIL). This approach decomposes representation learning into individual domains on a layer-by-layer basis and deeply couples learned multi-domain features through a dimension-staggered cascade. Drawing inspiration from physical principles, this architecture harmonizes separable extraction with cross-domain integration. While the orthogonality of the temporal, spatial, and frequency domains drives the separable learning for intra-domain structures, their shared origin as projections of the same transmission link dictates strong cross-domain correlation. We harness these properties through an interleaved multi-layer topology and inter-layer dimension-staggered cascade, fostering deep inter-domain synergy and feature enhancement.

These designs constrain the receptive field of each layer to a single physical dimension, resulting in superior parameter efficiency and computational parallelism. More importantly, by employing physical priors for perceptual compression, this FDIL architecture constrains learnable interactions to specific physical domains. This strategy circumvents the curse of increasing data dimensionality and introduces intrinsic regularization that mitigates overfitting in data-sparse and low signal-to-noise-ratio (SNR) environments. Its compact architecture also suggests considerable application potential for resource-constrained wireless devices. Building upon this versatile backbone, we demonstrate that classical learning operators—including one-dimensional (1D) convolutions, attention mechanisms, gating, and MLPs—can be easily instantiated into efficient channel representation networks with customizable functionalities. The effectiveness and superiority of these schemes are validated on the channel deduction task, a fundamental channel representation framework that encompasses estimation, interpolation, prediction, and feedback. The main contributions of this paper are summarized as follows:

- We formulate the multi-domain coupling priors of the time-space-frequency channel tensor and consequently propose a full-domain interleaved representation framework, providing a theoretical justification for moving beyond network stitching toward more cohesive wireless-native learning.
- We propose Coupler, a wireless-native channel representation backbone that uses three groups of individual modules to perform intra-domain mixing and guides inter-domain coupling through multi-module interleaving, achieving fine-grained collaborative learning across the time, space, and frequency domains.
- We develop a family of parameter-efficient model instances derived from Coupler, including the integration of phase-preserving complex-domain MLPs for spatial-frequency processing and the design of three lightweight temporal mechanisms based on convolutions, attention, and gated mixing, demonstrating the architecture's flexibility and versatility.
- We demonstrate the usage and superiority of Coupler and its derivatives on channel deduction, a fundamental representation learning task. Extensive experiments validate their significant accuracy gains, lightweight characteristics, and robust applicability to real-world measured data.

The remainder of this paper is organized as follows. Section 2 presents the problem formulation, including the system model, channel deduction framework, and limitations of existing models. Then, Section 3 analyzes the interlaced coupling property of the channel tensor and proposes Coupler based on FDIL to perform efficient channel deduction. Next, Section 4 shows the performance evaluation of various aspects of our proposed schemes. Finally, Section 5 draws the conclusion.

The following notations are used in this work: The fonts $\alpha$, $\mathbf{a}$, $\mathbf{H}$, and $\boldsymbol{\mathcal{H}}$ denote scalar, vector, matrix, and tensor variables, respectively. $\boldsymbol{\mathcal{H}}[l,m,n]$ denotes the $(l,m,n)$th element of tensor $\boldsymbol{\mathcal{H}}$. Since dimensional transformations of variables in neural networks are relatively complex, for clarity and conciseness, we consistently use uppercase boldface letters to denote matrix or tensor variables in neural-network operations when no ambiguity arises. In addition, $\hat{\mathbf{k}}$ denotes a unit direction vector, $\mathcal{A}$ denotes a parameter list or set, and $(\cdot)^{\mathrm{T}}$ and $(\cdot)^{\mathrm{H}}$ denote the transpose and Hermitian transpose operations, respectively. $\odot$ denotes the Hadamard product, and $\|\cdot\|_F$ denotes the Frobenius norm.

## 2. Problem formulation

In this section, we first introduce the channel model for MIMO-orthogonal frequency-division multiplexing (OFDM) systems and the channel deduction framework for continuous channel acquisition. We then review existing vanilla CDNets and further analyze their limitations.

### 2.1 System model

We consider a massive MIMO system, where a base station (BS) equipped with $N_{\mathrm{t}}$ antennas ($N_{\mathrm{t}} \gg 1$) serves multiple single-antenna users. The system employs OFDM modulation with $N_{\mathrm{c}}$ subcarriers. We first consider a static channel model and assume that the wireless channel between the BS and a user equipment (UE) consists of at most $P$ paths. The corresponding spatial-frequency channel matrix $\mathbf{H} \in \mathbb{C}^{N_{\mathrm{t}}\times N_{\mathrm{c}}}$ can be written as

$$\mathbf{H} = \sum_{p=1}^{P} \alpha_p \cdot e^{-j2\pi f_c \tau_p} \cdot \mathbf{a}(\hat{\mathbf{k}}_p) \cdot \mathbf{b}^{\mathrm{T}}(\tau_p) \tag{1}$$

where $f_c$ is the center frequency. For the $p$th path, $\alpha_p$, $\tau_p$, and $\hat{\mathbf{k}}_p$ represent the complex path gain, propagation delay, and three-dimensional (3D) unit vector of the departure direction, respectively. The array steering vector is defined as

$$\mathbf{a}(\hat{\mathbf{k}}) = \left[e^{j\frac{2\pi}{\lambda}\hat{\mathbf{k}}^{\mathrm{T}}\mathbf{d}_1}, e^{j\frac{2\pi}{\lambda}\hat{\mathbf{k}}^{\mathrm{T}}\mathbf{d}_2}, \cdots, e^{j\frac{2\pi}{\lambda}\hat{\mathbf{k}}^{\mathrm{T}}\mathbf{d}_{N_{\mathrm{t}}}}\right]^{\mathrm{T}} \tag{2}$$

where $\lambda$ is the carrier wavelength, and $\mathbf{d}_m$ denotes the position vector of the $m$th antenna with respect to the array center. The frequency-domain steering vector $\mathbf{b}(\tau)$ is defined as

$$\mathbf{b}(\tau) = \left[e^{-j2\pi\Delta f_1 \tau}, e^{-j2\pi\Delta f_2 \tau}, \cdots, e^{-j2\pi\Delta f_{N_c} \tau}\right]^{\mathrm{T}} \tag{3}$$

where $\Delta f_n$ is the frequency offset of the $n$th subcarrier relative to the center frequency $f_c$. According to the above channel model, for a given antenna array geometry and carrier configuration, the spatial-frequency channel matrix $\mathbf{H}$ is determined by the signal propagation path parameters between the BS and the UE. For notational simplicity, we denote this underlying physical parameter set as $\Theta = \{\alpha_p, \hat{\mathbf{k}}_p, \tau_p\}_{p=1}^{P}$.

We next consider the mobile-user scenario. Within one channel coherence time, the CSI can be considered approximately constant, whereas the previous CSI becomes outdated and unreliable over a coherence time. Therefore, practical wireless systems must continuously acquire CSI over consecutive time slots. We let $\mathbf{H}_t$ denote the channel at the $t$th time slot, which is inherently governed by the time-varying multi-path parameter set $\Theta_t = \{\alpha_{p,t}, \tau_{p,t}, \hat{\mathbf{k}}_{p,t}\}_{p=1}^{P_t}$. Over a consecutive time interval $[t_0 - N, \cdots, t_0 - 1, t_0]$, the channel matrices can be stacked along the temporal dimension to form a complex-valued 3D tensor:

$$\boldsymbol{\mathcal{H}} = \mathrm{Stack}(\mathbf{H}_{t_0-N}, \cdots, \mathbf{H}_{t_0-1}, \mathbf{H}_{t_0}) \in \mathbb{C}^{N'\times N_{\mathrm{t}}\times N_{\mathrm{c}}} \tag{4}$$

where $\mathrm{Stack}(\cdot)$ denotes the stacking operation along the time dimension, $t_0$ is the index of the current time slot, and $N' = N + 1$ is the total number of considered slots. This tensor provides a unified representation of mobile CSI by describing observations of the underlying electromagnetic propagation paths collected over multiple time slots, antennas, and subcarriers.

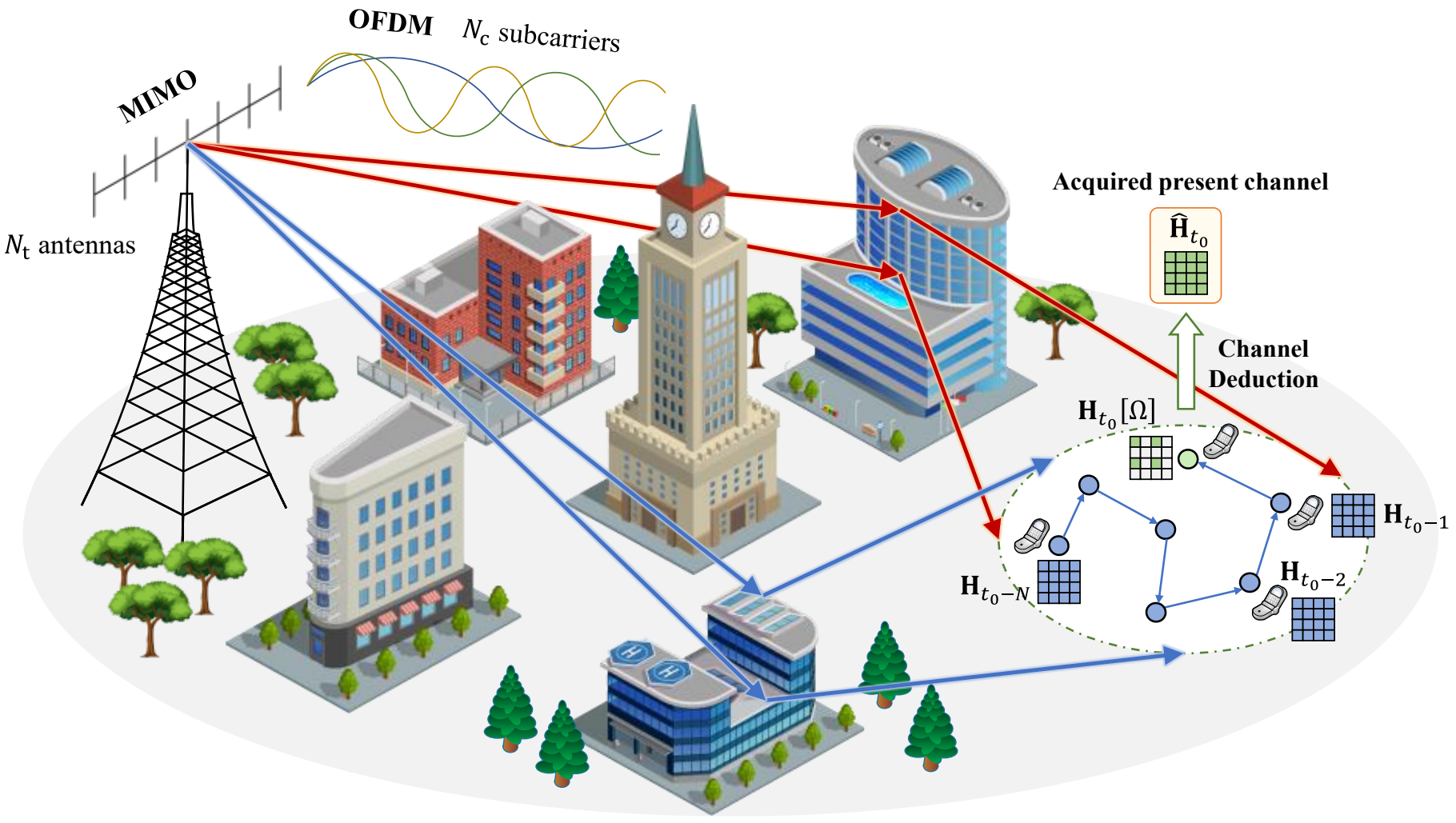

**Fig. 1.** Illustration of channel deduction for a mobile user. The current channel is deduced from past channels and the present sub-channel.

### 2.2 Channel deduction framework

In massive MIMO-OFDM systems, the channel is typically very high dimensional. To maintain communication efficiency, the signaling resources available for pilot transmission within each time slot are usually limited. This makes high-quality channel acquisition particularly challenging. To enable efficient CSI acquisition, it is necessary to fully exploit the correlations inherent in the channel across multiple physical domains, including time, space, and frequency.

First, according to Eq. (1), the channel responses of $\mathbf{H}_t$ over different antennas and subcarriers within a single time slot are governed by the same set of multi-path parameters, $\Theta_t$, which implies strong structural correlation within the spatial-frequency domain. Therefore, in practical systems, only a low-dimensional sub-channel $\mathbf{H}_t[\Omega] \in \mathbb{C}^{N_t^\Omega \times N_c^\Omega}$ is usually estimated in real time, and the complete channel $\mathbf{H}_t$ is then reconstructed by exploiting this spatial-frequency correlation. Here, $\Omega = \mathcal{A} \times \mathcal{B}$ denotes the two-dimensional (2D) indices of observed antennas and subcarriers, where $\mathcal{A}$ and $\mathcal{B}$ correspond to the subsets of $N_t^\Omega$ antennas and $N_c^\Omega$ subcarriers assigned for pilot transmission, respectively.

Second, over a sequence of consecutive time slots, the user movement is usually small relative to the scale of the macroscopic scattering environment. As a result, the multi-path structure of the channel generally does not change drastically. In other words, the path-parameter sets $\Theta_{t_0-N}, \ldots, \Theta_{t_0}$ of adjacent slots tend to remain similar at the large scale, which gives rise to implicit temporal correlation. Due to the potential randomness of user mobility and because the channel model in Eq. (1) contains a rapidly varying phase term $e^{-j2\pi f_c \tau_p}$ that is highly sensitive to small displacements, the channel in each time slot still exhibits its own transient characteristics.

Based on joint consideration of the intrinsic correlations of mobile channels in the temporal, spatial, and frequency domains, a channel deduction framework was proposed in Ref. [42]. This framework reconstructs the complete high-dimensional channel $\mathbf{H}_{t_0}$ at the current slot $t_0$ by leveraging both the previously acquired channel samples $\mathbf{H}_{t_0-N}, \cdots, \mathbf{H}_{t_0-1}$ from the past $N$ neighboring slots and the real-time coarse estimate $\mathbf{H}_{t_0}[\Omega]$. Formally, the process can be written as

$$\mathbf{H}_{t_0} = g_{\text{cd}}\left(\mathbf{H}_{t_0-N}, \cdots, \mathbf{H}_{t_0-1}, \mathbf{H}_{t_0}[\Omega]\right) \tag{5}$$

where $g_{\text{cd}}(\cdot)$ denotes the mapping function for channel deduction. Fundamentally, the channel deduction framework leverages the stability of large-scale multi-path structures across adjacent time slots, while simultaneously capturing unpredictable transient channel features through sparse pilots in the current slot, thereby enabling efficient and accurate channel acquisition.

Channel deduction can be viewed as a unified formulation for a broad class of wireless channel acquisition tasks, offering both strong generality and practical relevance. First, channel deduction can adapt to different system settings by reducing to several classical channel acquisition problems.

- **Reduction to channel mapping:** When temporal information is not used (i.e., $N = 0$), channel deduction degenerates into the single-slot channel mapping problem [25]. In this case, the system only uses the estimated $\mathbf{H}_{t_0}[\Omega]$ and spatial-frequency correlation to reconstruct the full channel $\mathbf{H}_{t_0}$.
- **Reduction to channel prediction:** When no pilots are transmitted in the current slot (i.e., $\mathbf{H}_{t_0}[\Omega]$ is unavailable), $g_{\text{cd}}(\cdot)$ extrapolates the current channel from historical channels. In this case, channel deduction then degenerates into channel prediction [16], [34].
- **Reduction to OFDM channel interpolation:** If the transmitter is equipped with only one antenna, as in certain uplink scenarios, $g_{\text{cd}}(\cdot)$ only needs to model temporal-frequency correlations. Therefore, it can be naturally applied to OFDM channel interpolation with temporal regression [44].

Channel deduction can also be extended to multiple practical channel acquisition applications.

- **CSI feedback in FDD systems:** Channel deduction also enables a UE-friendly CSI feedback scheme for FDD systems [45]. Instead of compressing and feeding back the full downlink CSI, $\mathbf{H}_{t_0}$, the UE only needs to report a partial observation $\mathbf{H}_{t_0}[\Omega]$. By combining the previously acquired downlink channels from past time slots, the BS can recover the full downlink CSI at the current slot via $g_{\text{cd}}(\cdot)$, thereby reducing the computational burden at the UE side.

- **Channel-database-assisted spatial channel deduction:** With the aid of a channel database, channel deduction can be extended from temporal channel acquisition to spatial channel deduction [43]. The BS can retrieve channel samples from the database within the spatial neighborhood of the target user and combine them with the current coarse observation to infer the user's full channel.
- **Scene-geometry-aided channel acquisition:** Channel deduction can also incorporate scene geometry information as environmental priors [46]. By leveraging geometric prompts such as coarse user location and building distribution, it can support environment-aware channel acquisition.

By subsuming multiple channel acquisition problems into a unified framework and systematically exploiting the multi-domain correlation structure of mobile channels, channel deduction can be regarded as a foundational task for channel representation learning. Consequently, the issue of how to realize the deduction function $g_{\mathrm{cd}}(\cdot)$ in an efficient and scalable manner becomes a fundamental problem for wireless channel representation learning.

### 2.3 DL-based channel deduction model

Due to the intricate coupling effects of multi-path propagation, the correlations of wireless channels across the temporal, spatial, and frequency domains are often implicit and highly nonlinear. For this reason, it is reasonable to design deep neural networks to learn the channel deduction function $g_{\mathrm{cd}}(\cdot)$ from data. Despite the feasibility of this channel acquisition framework, existing implementations predominantly rely on assembling conventional DL modules, making it difficult to achieve efficient multi-domain collaborative learning. For instance, vanilla CDNets [42] combine generic sequence models and existing CMixer modules [25] to perform temporal feature mixing and spatial-frequency channel mapping, respectively, as illustrated in Fig. 2. This "tool-driven" paradigm inherently fragments the channel representation process into decoupled stages, inevitably imposing profound architectural and structural limitations.

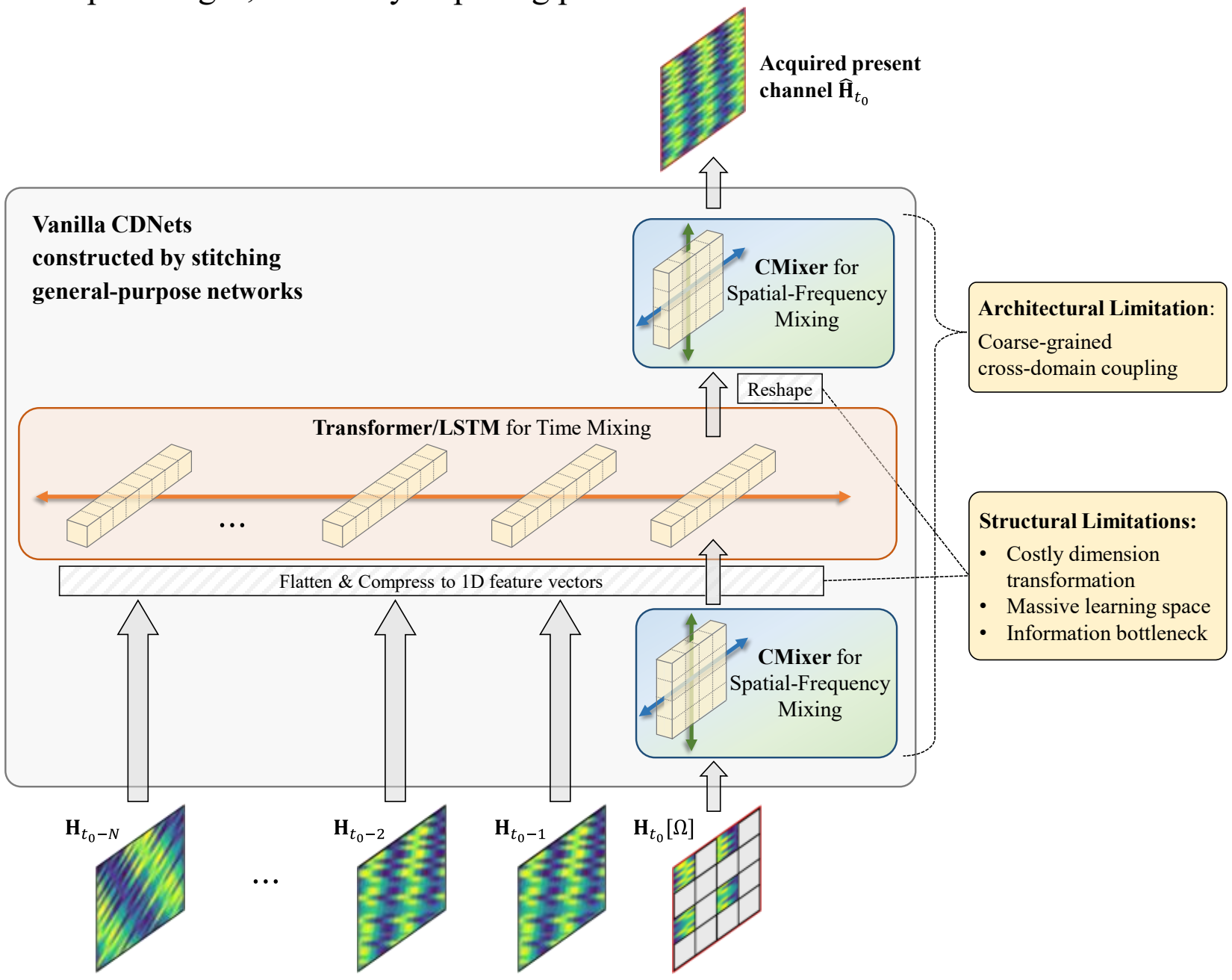

**Fig. 2**. The overall structure of vanilla CDNet [42], serving as a representative multi-stage decoupled architecture.

From an architectural perspective, such multi-stage designs suffer from coarse-grained cross-domain coupling. By strictly separating the feature extraction into a temporal phase and a spatial-frequency phase, the model only operates on the specific physical domain at each stage. Consequently, it cannot effectively interweave the dynamic temporal evolution with the spatial-frequency characteristics at a deep feature level and fails to fully exploit the intrinsic multi-domain coupling effect dictated by physical propagation laws.

From a structural perspective, adapting high-dimensional channel tensors to these off-the-shelf sequence modules necessitates costly channel reshaping. Classical sequence models operate on 1D feature vectors. To match this format, the 2D spatial-frequency channel matrices must be flattened and projected via a fully connected layer (e.g., $2N_{\mathrm{t}}N_{\mathrm{c}} \rightarrow S$). This operation not only destroys the intrinsic physical structure of the wireless channel but also results in a massive parameter space and severe information bottlenecks when the antenna and subcarrier dimensions are scaled up.

These limitations highlight the imperative need to transition from the conventional network stitching paradigm to a wireless-native backbone capable of enabling sufficient interaction within and across physical domains, maintaining a concise architecture that scales gracefully to large systems, and delivering superior parameter and computational efficiency.

## 3. Coupler: A full-domain interleaved learning backbone

In this section, we first reveal the unique multi-domain interlaced coupling property inherent in the temporal-spatial-frequency channel tensor $\boldsymbol{\mathcal{H}}$. Inspired by this property, we then develop the channel representation backbone Coupler via FDIL, along with its specific operator implementations. Next, we present a concise way to construct learning networks by backbone module stacking for the channel deduction task. Finally, we conduct complexity and property analyses of this neural backbone.

### 3.1 The interlaced coupling property of channel tensors

To develop an efficient and general-purpose backbone for channel representation learning, it is first necessary to examine the physical structure embodied in the temporal-spatial-frequency channel tensor $\boldsymbol{\mathcal{H}}$, which is the core object of representation learning. According to the channel model in Eq. (1) and channel tensor formulation in Eq. (4), each entry of this tensor can be expressed as

$$\boldsymbol{\mathcal{H}}[l,m,n] = \sum_{p=1}^{P_t} \alpha_{p,t} \cdot e^{-j2\pi(f_c+\Delta f_n)\tau_{p,t}} \cdot e^{j\frac{2\pi}{\lambda}\hat{\mathbf{k}}_{p,t}^{\mathrm{T}}\mathbf{d}_m} = q(\Theta_t, \mathbf{d}_m, \Delta f_n) \tag{6}$$

where the temporal index $l$ of the tensor and the time slot index $t$ satisfy $l = t - t_0 + N + 1$, and the channel structure function $q(\cdot,\cdot,\cdot)$ uniformly describes the coupled impact of the multi-path parameters, antenna positions, and subcarrier frequencies on the channel response. Furthermore, we define the feature parameter lists for the temporal, spatial, and frequency domains as $\mathcal{T} = [\Theta_{t_0-N}, \cdots, \Theta_{t_0}]$, $\mathcal{S} = [\mathbf{d}_1, \cdots, \mathbf{d}_{N_\mathrm{t}}]$, and $\mathcal{F} = [\Delta f_1, \cdots, \Delta f_{N_\mathrm{c}}]$, respectively. Consequently, the channel tensor can be compactly formulated as

$$\boldsymbol{\mathcal{H}} = q(\mathcal{T} \times \mathcal{S} \times \mathcal{F}) \tag{7}$$

where $\mathcal{T} \times \mathcal{S} \times \mathcal{F}$ denotes the 3D parameter grid generated by these three lists. This expression profoundly reveals the intrinsic nature of the high-dimensional channel tensor $\boldsymbol{\mathcal{H}}$: It is fundamentally formed by the *interlaced* coupling of the temporal, spatial, and frequency feature parameters through the shared channel structure function $q(\cdot,\cdot,\cdot)$. Correspondingly, the correlations along the temporal, spatial, and frequency domains are primarily governed by the parameter lists $\mathcal{T}$, $\mathcal{S}$, and $\mathcal{F}$, respectively. Taking the temporal domain as an example, for any given antenna index $m$ and subcarrier index $n$, the temporal feature vector can be denoted as $\boldsymbol{\mathcal{H}}[:,m,n] = q(\mathcal{T}, \mathbf{d}_m, \Delta f_n)$. This indicates that temporal vectors sliced at different antenna and subcarrier indices share the same time-varying multi-path parameter list $\mathcal{T} = [\Theta_{t_0-N}, \cdots, \Theta_{t_0}]$, as well as the same underlying structure function $q(\cdot,\cdot,\cdot)$, differing only in the coupled spatial feature $\mathbf{d}_m$ and frequency feature $\Delta f_n$. Therefore, these temporal vectors consistently reflect the time evolution of the channel. By the same reasoning, the spatial slice vectors $\{\boldsymbol{\mathcal{H}}[l,:,n]\}_{l,n}$ primarily characterize the shared antenna array properties $\mathcal{S}$, whereas the frequency slice vectors $\{\boldsymbol{\mathcal{H}}[l,m,:]\}_{l,m}$ primarily characterize the shared frequency properties $\mathcal{F}$.

In summary, the internal correlation of the complete CSI tensor fundamentally arises from the interlaced coupling among the temporal feature $\mathcal{T}$, spatial feature $\mathcal{S}$, and frequency feature $\mathcal{F}$ under the shared channel structure function $q(\cdot,\cdot,\cdot)$. Moreover, although the tensor $\boldsymbol{\mathcal{H}}$ has a data dimensionality on the order of $O(N'N_\mathrm{t}N_\mathrm{c})$, the intrinsic feature dimensionality that determines its structure is only $O(N' + N_\mathrm{t} + N_\mathrm{c})$. This indicates that, despite its high dimensionality, the channel tensor possesses a limited number of intrinsic degrees of freedom; hence, its representation learning process can be substantially simplified by a properly designed neural architecture.

### 3.2 Full-domain interleaved learning architecture

As revealed in Section 3.1, a high-dimensional channel tensor is intrinsically formed by the interlaced coupling of low-dimensional temporal, spatial, and frequency features. Driven by this physical insight, we recognize that the representation learning of the channel tensor does not require simultaneous processing across all dimensions. Instead, it can be naturally decomposed into successive operations over individual physical domains, effectively avoiding a prohibitively large parameter space and excessive learning burden.

Building upon this principle, we propose an FDIL architecture constructed by stacking $K$ identical blocks. Each block alternately performs feature mixing along the temporal, spatial, and frequency dimensions of the channel tensor $\boldsymbol{\mathcal{H}}$, thereby fully exploiting the intra-domain correlations within each physical domain while conducting structural interlaced coupling across different domains. For the $k$th block, the computation is given by the following:

$$\mathbf{X}^{(k)} = \mathbf{Z}^{(k-1)} + \mathrm{TimeMixing}(\mathrm{LN}(\mathbf{Z}^{(k-1)})) \tag{8}$$

$$\mathbf{Y}^{(k)} = \mathbf{X}^{(k)} + \mathrm{SpaceMixing}(\mathrm{LN}(\mathbf{X}^{(k)})) \tag{9}$$

$$\mathbf{Z}^{(k)} = \mathbf{Y}^{(k)} + \mathrm{FreqMixing}(\mathrm{LN}(\mathbf{Y}^{(k)})) \tag{10}$$

where $k = 1, \cdots, K$, and the model input $\mathbf{Z}^{(0)} = \boldsymbol{\mathcal{H}}$. Additionally, TimeMixing, SpaceMixing, and FreqMixing are computed along the temporal, frequency, and spatial dimensions of each layer's input tensor, respectively, thereby forming a *dimension-staggered* multi-module cascaded computation flow within Eq. (8)–(10). Residual connections and layer normalization (LN) are used to ensure training stability in deep networks. Unlike the conventional paradigm of assembling generic network modules to process multiple physical domains in separate stages, this novel architecture inherently enables fine-grained cross-domain coupling. Specifically, when multiple coupler blocks are stacked to construct the model backbone, the overall architecture exhibits *interleaved* learning patterns across time, space, frequency, time, and so forth, in a cyclical manner. Through this layer-by-layer alternate mixing, the channel features learned across multiple physical domains are deeply interwoven in a manner well aligned with the *interlaced* coupling structure of CSI revealed in Section 3.1. After $K$ FDIL blocks, the output representation tensor $\mathbf{Z}^{(K)}$ fully integrates the channel features across all physical domains. Since this learning architecture efficiently couples feature interactions across all physical domains within a unified framework, we term the resulting backbone network "Coupler."

Next, we detail the design of the mixing modules in Coupler by considering the distinct channel properties in different physical domains. In the spatial and frequency domains, the channel responses can be viewed as observations of the same set of multi-path parameters over different antenna locations and subcarrier frequencies. Their structures are mainly determined by the array geometry and the frequency allocation, which are typically fixed system configurations. As a result, these two domains exhibit relatively static structural patterns with fixed dimensionality. Based on this property, we adopt a complex-domain multilayer perceptron (CMLP) for feature mixing in both the spatial and frequency domains. In the spatial domain, a CMLP with input/output dimension $2N_\mathrm{t}$ and hidden width $S_\mathrm{t}$ is used for spatial mixing, and its parameters are shared across all $N'$ time slots and $N_\mathrm{c}$ subcarriers. Similarly, in the frequency domain, a CMLP with input/output dimension $2N_\mathrm{c}$ and hidden width $S_\mathrm{c}$ is employed, with the same operation shared across all time slots and antennas. As illustrated in the upper-right corner of Fig. 3, the CMLP jointly processes the real and imaginary parts of each feature vector, thereby preserving the integrity of the complex-domain information during feature extraction.

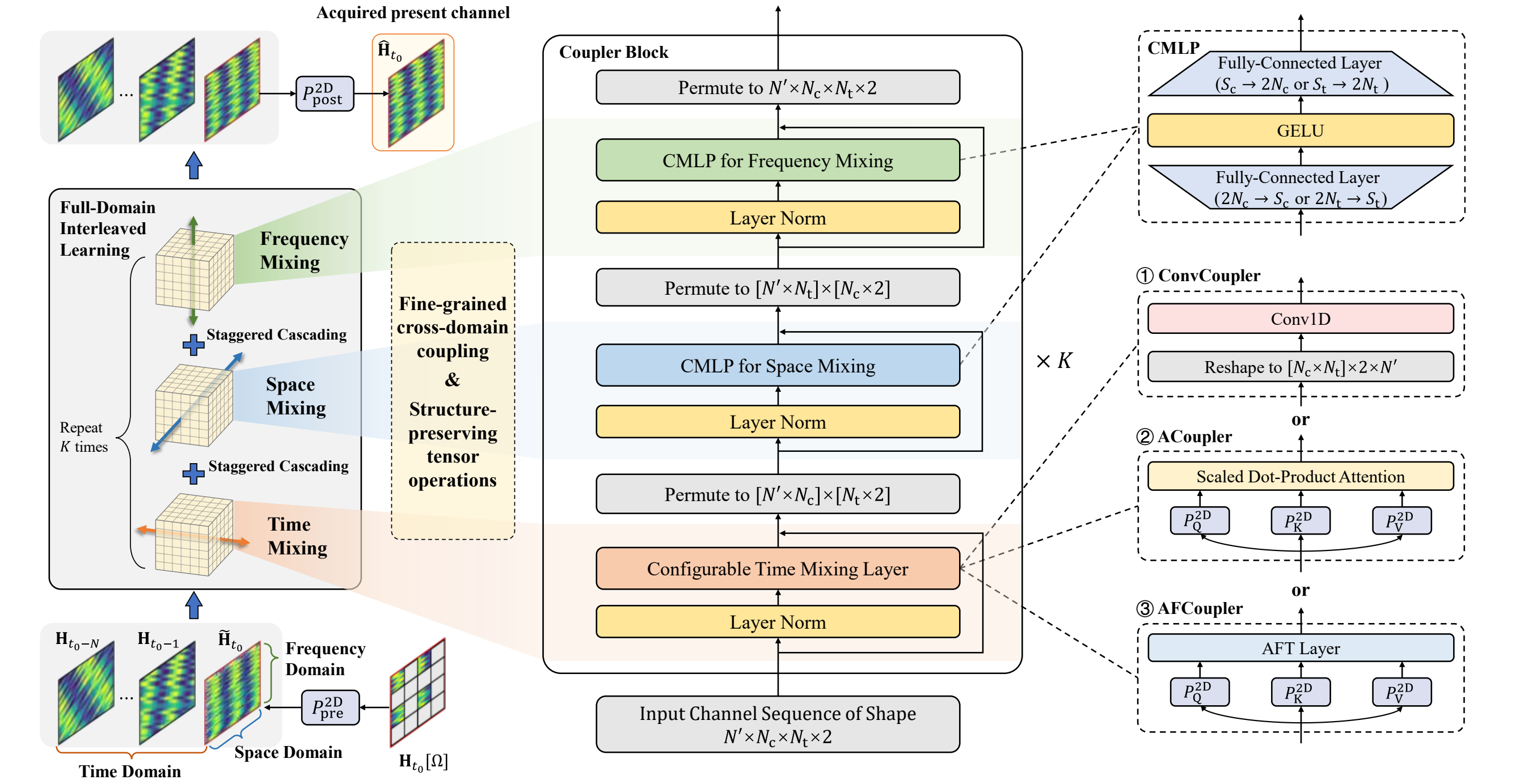

**Fig. 3.** Overall architecture of the Coupler backbone, consisting of cascaded temporal, spatial, and frequency feature mixing modules. The spatial and frequency mixing modules are implemented using CMLPs, while the temporal mixing module admits three different realizations, corresponding to ConvCoupler, ACoupler, and AFCoupler, respectively. With the interleaved connected (IC)-2D layers $P_{\mathrm{pre}}^{\mathrm{2D}}$ and $P_{\mathrm{post}}^{\mathrm{2D}}$, Coupler can be readily applied to channel deduction.

In contrast, temporal feature learning is substantially more challenging. First, unlike the spatial and frequency dimensions, the temporal length $N'$ of the channel tensor is variable. Second, the temporal characteristics of the channel are governed by the time-evolving multi-path parameter sequence $[\Theta_{t_0-N}, \dots, \Theta_{t_0}]$, whose variation is more complex and does not exhibit fixed patterns like those arising from antenna-array geometry or frequency-band allocation. Therefore, CMLPs are no longer suitable for temporal modeling. To address this issue, we design three temporal mixing modules with different modeling capabilities and inductive biases.

### *3.2.1 ConvCoupler*

We first adopt 1D convolution, as the simplest temporal information mixing mechanism, to construct a convolution-based Coupler (ConvCoupler). Specifically, the channel tensor is reshaped into $N_{\mathrm{t}}N_{\mathrm{c}} \times 2 \times N'$, and a 1D convolution with kernel size $K_{\mathrm{conv}}$ and two input/output channels is applied to process the temporal sequence of length $N'$, where padding is used to keep the temporal dimension unchanged. This process can be written as

$$\begin{aligned} &\mathbb{R}^{N'\times N_{\mathrm{t}}\times N_{\mathrm{c}}\times 2} \left(\mathbb{C}^{N'\times N_{\mathrm{t}}\times N_{\mathrm{c}}}\right) \xrightarrow{\text{reshape}} \mathbb{R}^{N_{\mathrm{t}}N_{\mathrm{c}}\times 2\times N'} \\ &\xrightarrow{\text{Conv1D}} \mathbb{R}^{N_{\mathrm{t}}N_{\mathrm{c}}\times 2\times N'} \xrightarrow{\text{reshape}} \mathbb{R}^{N'\times N_{\mathrm{t}}\times N_{\mathrm{c}}\times 2} \left(\mathbb{C}^{N'\times N_{\mathrm{t}}\times N_{\mathrm{c}}}\right) \end{aligned} \tag{11}$$

It can be seen that, for each spatial-frequency position in the channel feature tensor, ConvCoupler uses the same temporal mixing weights. In other words, it naturally extends the parameter-sharing principle used in spatial and frequency mixing—namely, sharing parameters across the physical dimensions other than the one currently being mixed—to the temporal modeling process. Although this temporal mixing mechanism is relatively simple, it can still work effectively when deeply stacked under the FDIL architecture of Coupler.

### *3.2.2 ACoupler*

In ConvCoupler, the temporal operation uses fixed connection weights along the temporal dimension and thus constitutes a static-mixing mechanism. Due to the local receptive-field property of convolution kernels, it mainly captures local correlations between neighboring time slots. To enhance the model's ability to capture long-range dependencies, we further introduce a self-attention mechanism to capture correlations between arbitrary pairs of time slots.

However, standard attention operates on 1D token sequences. If directly applied to the channel tensor, the 2D spatial-frequency channel at each time slot must first be flattened and compressed into a 1D feature, which leads to two problems: It destroys the original 2D structure of the spatial-frequency channel, and it incurs prohibitively large parameter and computational costs for high-dimensional channels. To address this issue, we design a complex-domain interleaved connected 2D layer (IC-2D) to generate the query, key, and value representations in self-attention. In the general case where the input is $\mathbb{C}^{N_1\times N_2}$ and the output is $\mathbb{C}^{N_1'\times N_2'}$, the computation of the IC-2D layer $P^{\mathrm{2D}}(\cdot)$ is

$$\begin{aligned} &\mathbb{R}^{N_1\times N_2\times 2} \left(\mathbb{C}^{N_1\times N_2}\right) \xrightarrow{\text{reshape}} \mathbb{R}^{N_2\times 2N_1} \xrightarrow{\text{FC}_1} \mathbb{R}^{N_2\times 2N_1'} \\ &\xrightarrow{\text{reshape}} \mathbb{R}^{N_1'\times 2N_2} \xrightarrow{\text{FC}_2} \mathbb{R}^{N_1'\times 2N_2'} \xrightarrow{\text{reshape}} \mathbb{R}^{N_1'\times N_2'\times 2} \left(\mathbb{C}^{N_1'\times N_2'}\right) \end{aligned} \tag{12}$$

where $FC_1$ and $FC_2$ denote the fully connected layers for the $2N_1 \to 2N_1'$ and $2N_2 \to 2N_2'$ transformations, respectively. Using IC-2D, we project the normalized spatial-frequency token at each time slot, denoted by $\bar{\mathbf{Z}}_l^{(k-1)} \in \mathbb{R}^{N_\mathrm{t} \times N_\mathrm{c} \times 2}$, into query, key, and value tensors of the same size:

$$\mathbf{Q}_l^{(k)} = P_\mathrm{Q}^{2\mathrm{D}}(\bar{\mathbf{Z}}_l^{(k-1)}), \quad \mathbf{K}_l^{(k)} = P_\mathrm{K}^{2\mathrm{D}}(\bar{\mathbf{Z}}_l^{(k-1)}), \quad \mathbf{V}_l^{(k)} = P_\mathrm{V}^{2\mathrm{D}}(\bar{\mathbf{Z}}_l^{(k-1)}) \tag{13}$$

where $\bar{\mathbf{Z}}_l^{(k-1)} = \mathrm{LN}(\mathbf{Z}_l^{(k-1)})$, and the subscript $l$ denotes the slice at the $l$th time slot. Compared with flattened fully connected projections, IC-2D significantly reduces the number of parameters while preserving the 2D spatial-frequency structure, avoiding the information loss potentially caused by dimensional compression.

We then perform temporal interaction via scaled dot-product attention (SDPA) [47], treating the combined spatial-frequency and complex-valued components as the feature representation for each temporal token:

$$\mathrm{SDPA}(\mathbf{Q}, \mathbf{K}, \mathbf{V}) = \mathrm{softmax}\left(\frac{\mathbf{Q}\mathbf{K}^\mathrm{T}}{\sqrt{2N_\mathrm{t}N_\mathrm{c}}}\right)\mathbf{V} \tag{14}$$

Here, for notational simplicity, the block superscript $(k)$ is omitted when no ambiguity arises, and $\mathbf{Q}$, $\mathbf{K}$, and $\mathbf{V}$ are reshaped from $N' \times N_\mathrm{t} \times N_\mathrm{c} \times 2$ into $N' \times 2N_\mathrm{t}N_\mathrm{c}$ before the attention operation. Through SDPA, the resulting attention-based Coupler (ACoupler) is able to capture global correlations between arbitrary time slots.

### *3.2.3 AFCoupler*

According to the experimental results in Ref. [42], the attention-based CDNet (ACDNet) generally attains a higher performance ceiling than the recurrent-based CDNet (RCDNet) but is less robust to errors in the input channels. This is because ACDNet tends to infer the current channel by exploiting small-scale feature correlations across time slots to utilize the most similar historical channels. Although this mechanism can improve performance under ideal conditions, its reliance on small-scale features makes it more sensitive to input errors. In contrast, RCDNet, through recursive feature accumulation, is better at extracting the shared large-scale characteristics across all time slots and therefore offers greater robustness. However, due to the sequential nature of RNNs, it is difficult to parallelize and thus not well suited to the Coupler architecture. For this reason, we seek a temporal mixing mechanism with a full temporal receptive field, stronger learning capability for stable global features, and compatibility with parallel computation.

Inspired by the linear-attention mechanism in the attention-free Transformer (AFT) [48], we propose the attention-free Coupler (AFCoupler). Specifically, AFCoupler first generates query $\mathbf{Q}$, key $\mathbf{K}$, and value $\mathbf{V}$ using the IC-2D projection and then performs temporal mixing through an AFT layer:

$$[\mathrm{AFT}(\mathbf{Q}, \mathbf{K}, \mathbf{V})]_l = \sigma(\mathbf{Q}_l) \odot \sum_{l'=1}^{N'} (\mathrm{softmax}(\mathbf{K}) \odot \mathbf{V})_{l'} = \sigma(\mathbf{Q}_l) \odot \frac{\sum_{l'=1}^{N'} \exp(\mathbf{K}_{l'}) \odot \mathbf{V}_{l'}}{\sum_{l'=1}^{N'} \exp(\mathbf{K}_{l'})} \tag{15}$$

where $\sigma(\cdot)$ denotes the sigmoid function. Mechanistically, the AFT layer first computes a global key-value summary from $\mathbf{K}$ and $\mathbf{V}$ to characterize the shared large-scale multi-path structure in $\{\mathbf{H}_{t_0-N}, \dots, \mathbf{H}_{t_0}\}$. This shared structure arises from the common scattering environment and the spatial proximity of the UE across adjacent time slots. The query $\mathbf{Q}$ generated from each spatial-frequency token acts as a gating signal to embed the slot-specific spatial-frequency characteristics, thereby capturing transient channel variations caused by user mobility randomness and rapid phase changes. As a result, AFCoupler can not only learn stable global large-scale features more effectively but also retain the ability to characterize local small-scale variations at each time slot.

In summary, by alternately performing temporal, spatial, and frequency feature mixing, we propose Coupler, a backbone architecture based on FDIL, for efficient channel-tensor representation learning. It transforms a complex-valued channel tensor of size $N' \times N_\mathrm{t} \times N_\mathrm{c}$ to a feature tensor of the same dimension. By incorporating different temporal mixing mechanisms, we obtain three Coupler derivatives with distinct temporal modeling characteristics; their performance in different cases will be comparatively evaluated in the subsequent experiments. Fig. 3 illustrates the overall architecture of Coupler, together with the structural designs of its spatial-frequency and temporal mixing modules.

## 3.3 Channel deduction based on Coupler

Based on the Coupler backbone, a complete channel deduction model can be constructed by introducing only a few simple dimensional transformation modules. First, an IC-2D layer with the mapping $\mathbb{C}^{N_\mathrm{t}^\Omega \times N_\mathrm{c}^\Omega} \to \mathbb{C}^{N_\mathrm{t} \times N_\mathrm{c}}$ is employed to pre-map the observed sub-channel at the current time slot into the full channel dimension:

$$\tilde{\mathbf{H}}_{t_0} = P_\mathrm{pre}^{2\mathrm{D}}\big(\mathbf{H}_{t_0}[\Omega]\big) \in \mathbb{C}^{N_\mathrm{t} \times N_\mathrm{c}} \tag{16}$$

As discussed previously, this pre-mapping is both parameter-efficient and consistent with the inherent 2D structure of the spatial-frequency channel. The pre-mapped current channel is then stacked with the channel sequence from previous time slots along the temporal dimension to form the input tensor of Coupler:

$$\mathbf{Z}^{(0)} = \mathrm{Stack}\big(\mathbf{H}_{t_0-N}, \cdots, \mathbf{H}_{t_0-1}, \tilde{\mathbf{H}}_{t_0}\big) \tag{17}$$

After passing through $K$ Coupler blocks, the channel information from the current and historical slots is jointly fused across the temporal, spatial, and frequency domains in a unified and efficient manner. Finally, the spatial-frequency feature map $\mathbf{Z}_{N'}^{(K)}$ corresponding to the current time slot is extracted from the output tensor $\mathbf{Z}^{(K)}$ and fed into an IC-2D output head for lightweight post-processing, yielding the reconstructed channel at the current slot:

$$\hat{\mathbf{H}}_{t_0} = P_\mathrm{post}^{2\mathrm{D}}\left(\mathbf{Z}_{N'}^{(K)}\right) \tag{18}$$

This completes the channel deduction process. The computational flow of the channel deduction model based on the Coupler architecture is illustrated on the left side of Fig. 3. In the following, we use the term "Couplers" to collectively refer to the channel deduction models proposed in this paper.

For training, continuous channel sequences naturally arising in the communication process can be directly used to construct training samples for Couplers, making it easy to collect the required training data. Moreover, because the channel samples within a short-term mobile sequence $\{\mathbf{H}_{t_0-N}, \dots, \mathbf{H}_{t_0}\}$ remain in a finite neighborhood, additional valid training samples can be generated by resampling and reordering subsequences from the collected data, thereby improving data utilization and enhancing generalization [42]. The model is optimized using the mean squared error (MSE) loss:

$$\mathcal{L} = \frac{1}{num} \sum_{i=1}^{num} \left[ \left\| \widehat{\mathbf{H}}_{t_0} - \mathbf{H}_{t_0} \right\|_F^2 \right]_i \tag{19}$$

where $num$ denotes the number of training samples.

After the Couplers have been trained, we obtain a usable function $g_{\text{cd}}(\cdot)$ that can deduce $\mathbf{H}_{t_0}$ based on $\mathbf{H}_{t_0-N}, \dots, \mathbf{H}_{t_0-1}, \mathbf{H}_{t_0}[\Omega]$. This deduction process can be continued in an autoregressive manner; that is, the deduced channel $\widehat{\mathbf{H}}_t$ can be reused as one of the historical inputs for inferring $\mathbf{H}_{t+1}$. In this way, it is only necessary to use high-density pilots to obtain $\mathbf{H}_0, \dots, \mathbf{H}_{N-1}$ in the initial access stage, and the channels $\mathbf{H}_N, \mathbf{H}_{N+1}, \dots$ of subsequent slots can then be continuously inferred through autoregressive channel deduction using the sub-channel $\mathbf{H}_N[\Omega], \mathbf{H}_{N+1}[\Omega], \dots$ estimated from sparse pilots.

It is worth noting that the Coupler backbone is not restricted to channel deduction. Since its design exploits the intrinsic structure of spatial-temporal-frequency channel tensors, Coupler can be extended to broader wireless channel representation learning tasks by adjusting the input format and task-specific head. For example, it can be used to learn environment-aware representations from multiple complete spatial-frequency channels, $\mathbf{H}_1, \mathbf{H}_2, \dots, \mathbf{H}_N$, supporting tasks such as channel generation, user localization, and scenario sensing. It can also process sparse-pilot-based channel observations, $\mathbf{H}_1[\Omega_1], \mathbf{H}_2[\Omega_2], \dots, \mathbf{H}_N[\Omega_N]$, for more challenging channel acquisition tasks such as multi-user joint channel estimation.

### 3.4 Complexity and property analysis

In this sub-section, we systematically compare the proposed Couplers with the existing CDNets and analyze the advantages of Couplers in terms of parameter efficiency and architectural generality. CDNets adopt a cascaded structure consisting of a spatial-frequency mixing module and a temporal mixing module. The spatial-frequency mixing module has a parameter count of $\mathcal{O}(N_\text{t}S_\text{t} + N_\text{c}S_\text{c})$ and a computational complexity of $\mathcal{O}\big(N_\text{t}N_\text{c}(S_\text{t} + S_\text{c})\big)$. Before temporal modeling, CDNets must flatten the 2D spatial-frequency channel and project it into a 1D feature of size $S$, which introduces $\mathcal{O}(N_\text{t}N_\text{c}S)$ parameters and $\mathcal{O}(N'N_\text{t}N_\text{c}S)$ computational complexity. Subsequently, RCDNet performs temporal modeling using an LSTM, resulting in $\mathcal{O}(S^2)$ parameters and $\mathcal{O}(N'S^2)$ computational complexity. ACDNet performs temporal modeling using a Transformer, corresponding to $\mathcal{O}(S^2)$ parameters and $\mathcal{O}(N'S^2 + (N')^2S)$ computational complexity.

In Couplers, each Coupler block embeds temporal, spatial, and frequency feature mixing operations in an alternating manner. For ConvCoupler, the temporal mixing layer has $\mathcal{O}(K_\text{conv})$ parameters and $\mathcal{O}(N'N_\text{t}N_\text{c}K_\text{conv})$ operations. For both ACoupler and AFCoupler, the parameters of the temporal mixing layer come from the IC-2D projections, amounting to $\mathcal{O}(N_\text{t}^2 + N_\text{c}^2)$. Their computational complexities are $\mathcal{O}(N'^2N_\text{t}N_\text{c} + N'N_\text{t}^2N_\text{c} + N'N_\text{t}N_\text{c}^2)$ and $\mathcal{O}(N'N_\text{t}^2N_\text{c} + N'N_\text{t}N_\text{c}^2)$, respectively. Across all variants, the CMLP-based spatial and frequency mixing layers introduce $\mathcal{O}(N_\text{t}S_\text{t} + N_\text{c}S_\text{c})$ parameters and require $\mathcal{O}\big(N'N_\text{t}N_\text{c}(S_\text{t} + S_\text{c})\big)$ operations.

It is worth noting that the spatial hidden width $S_\text{t}$, the frequency hidden width $S_\text{c}$, and the spatial-frequency feature dimension $S$ are usually set to be on the same order as their corresponding physical channel dimensions (i.e., $S_\text{t} \propto N_\text{t}$, $S_\text{c} \propto N_\text{c}$, and $S \propto N_\text{t}N_\text{c}$), so that the learned features can characterize the physical properties along the corresponding dimensions without dimensionality reduction. Under this model-parameter setting, Table 1 summarizes the simplified asymptotic parameter counts and floating-point operations (FLOPs) of CDNets and Couplers.

**Table 1**
Simplified asymptotic parameter counts and FLOPs under $S_\text{t} = \mathcal{O}(N_\text{t})$, $S_\text{c} = \mathcal{O}(N_\text{c})$, and $S = \mathcal{O}(N_\text{t}N_\text{c})$.

| Model | Parameter counts | | FLOPs | |
|---|---|---|---|---|
| | TM | SM & FM | TM | SM & FM |
| ConvCoupler | $\mathcal{O}(K_\text{conv})$ | $\mathcal{O}(N_\text{t}^2 + N_\text{c}^2)$ | $\mathcal{O}(N'N_\text{t}N_\text{c}K_\text{conv})$ | $\mathcal{O}\big(N'N_\text{t}N_\text{c}(N_\text{t} + N_\text{c})\big)$ |
| ACoupler | $\mathcal{O}(N_\text{t}^2 + N_\text{c}^2)$ | $\mathcal{O}(N_\text{t}^2 + N_\text{c}^2)$ | $\mathcal{O}\big(N'N_\text{t}N_\text{c}(N' + N_\text{t} + N_\text{c})\big)$ | $\mathcal{O}\big(N'N_\text{t}N_\text{c}(N_\text{t} + N_\text{c})\big)$ |
| AFCoupler | $\mathcal{O}(N_\text{t}^2 + N_\text{c}^2)$ | $\mathcal{O}(N_\text{t}^2 + N_\text{c}^2)$ | $\mathcal{O}\big(N'N_\text{t}N_\text{c}(N_\text{t} + N_\text{c})\big)$ | $\mathcal{O}\big(N'N_\text{t}N_\text{c}(N_\text{t} + N_\text{c})\big)$ |
| RCDNet | $\mathcal{O}(N_\text{t}^2N_\text{c}^2)$ | $\mathcal{O}(N_\text{t}^2 + N_\text{c}^2)$ | $\mathcal{O}(N'N_\text{t}^2N_\text{c}^2)$ | $\mathcal{O}\big(N_\text{t}N_\text{c}(N_\text{t} + N_\text{c})\big)$ |
| ACDNet | $\mathcal{O}(N_\text{t}^2N_\text{c}^2)$ | $\mathcal{O}(N_\text{t}^2 + N_\text{c}^2)$ | $\mathcal{O}((N')^2N_\text{t}N_\text{c} + N'N_\text{t}^2N_\text{c}^2)$ | $\mathcal{O}\big(N_\text{t}N_\text{c}(N_\text{t} + N_\text{c})\big)$ |

Note: TM, temporal mixing; SM, spatial mixing; FM, frequency mixing.

In terms of parameter efficiency, the FDIL architecture of Couplers preserves all intermediate features in the tensor form $N' \times N_\text{t} \times N_\text{c} \times 2$, instead of flattening the 2D spatial-frequency channel into 1D tokens. By assigning learnable parameters only along one physical dimension at a time, Couplers effectively avoid a large parameter space. As a result, they substantially reduce the parameter burden of temporal mixing compared with CDNets, leading to higher overall parameter and learning efficiency.

In terms of computational complexity, when the spatial-frequency channel dimensions are relatively small and the spatial-frequency feature size $S$ in CDNets is set to a moderate fixed value, Couplers may exhibit slightly higher FLOPs because they perform spatial-frequency feature mixing in parallel across all time slots. However, the temporal mixing layers of Couplers do not require the 2D spatial-frequency channel to be flattened, and the overall computational complexity can be controlled within $O\left(N'N_{\mathrm{t}}N_{\mathrm{c}}(N'+N_{\mathrm{t}}+N_{\mathrm{c}})\right)$. In contrast, CDNets perform temporal modeling on flattened $S$-size spatial-frequency features, resulting in a temporal computational complexity of $O(N'N_{\mathrm{t}}^2N_{\mathrm{c}}^2)$. Therefore, for high-dimensional channels, Couplers exhibit better scalability in both parameter count and computational complexity, while achieving a more balanced allocation of model parameters and computations across different physical dimensions. Such a compact architecture with lightweight operators and parallel-friendly computation is potentially beneficial for model deployment on resource-constrained wireless devices and certain dedicated hardware platforms.

Beyond efficiency, Coupler serves as a generalized framework for channel representation learning that unifies several classical architectures. For instance, if the temporal-domain mixing is omitted, Coupler reduces to the CMixer network for space-frequency channel mapping [25]. Similarly, in single-antenna (omitting the spatial mixing) or single-carrier (omitting the frequency mixing) scenarios, Coupler equivalently reduces to classic sequential learning architectures that capture the temporal evolution of 1D frequency or spatial features, driven by sequence models such as Transformer [47] or AFT [48]. Crucially, this architecture ensures that expanding the learning capacity remains a linear process—simply incorporating an additional operator for each new physical domain. This effectively bypasses the combinatorial explosion of parameters seen in conventional models, where complexity scales with the product of input dimensions, thus significantly mitigating structural redundancy.

## 4. Numerical experiments

In this section, we comprehensively evaluate the performance and properties of the proposed Couplers. First, we introduce the experimental settings. Then, we assess the Couplers' performance on the channel deduction task in terms of channel acquisition accuracy, impact of the number of available past channels, robustness to lossy inputs, effect of training data size, and autoregressive deduction under error propagation. Finally, we further verify the effectiveness and application value of Coupler-based channel deduction on real-world data collected from a practical wireless system.

### 4.1 Experimental settings

#### *4.1.1 Communication scenario*

In this work, we first conduct simulation experiments using the raytracing-based DeepMIMO dataset [49]. Specifically, we generate wireless channel data in a typical outdoor communication scenario "O1," as shown in Fig. 4, where a BS equipped with a uniform linear array (ULA) provides communication services to users located within the red box area. In this scenario, we assume that many users maintain communication connections with the BS. By leveraging user movement within the service area, we can collect extensive channel data from diverse locations to train and test models.

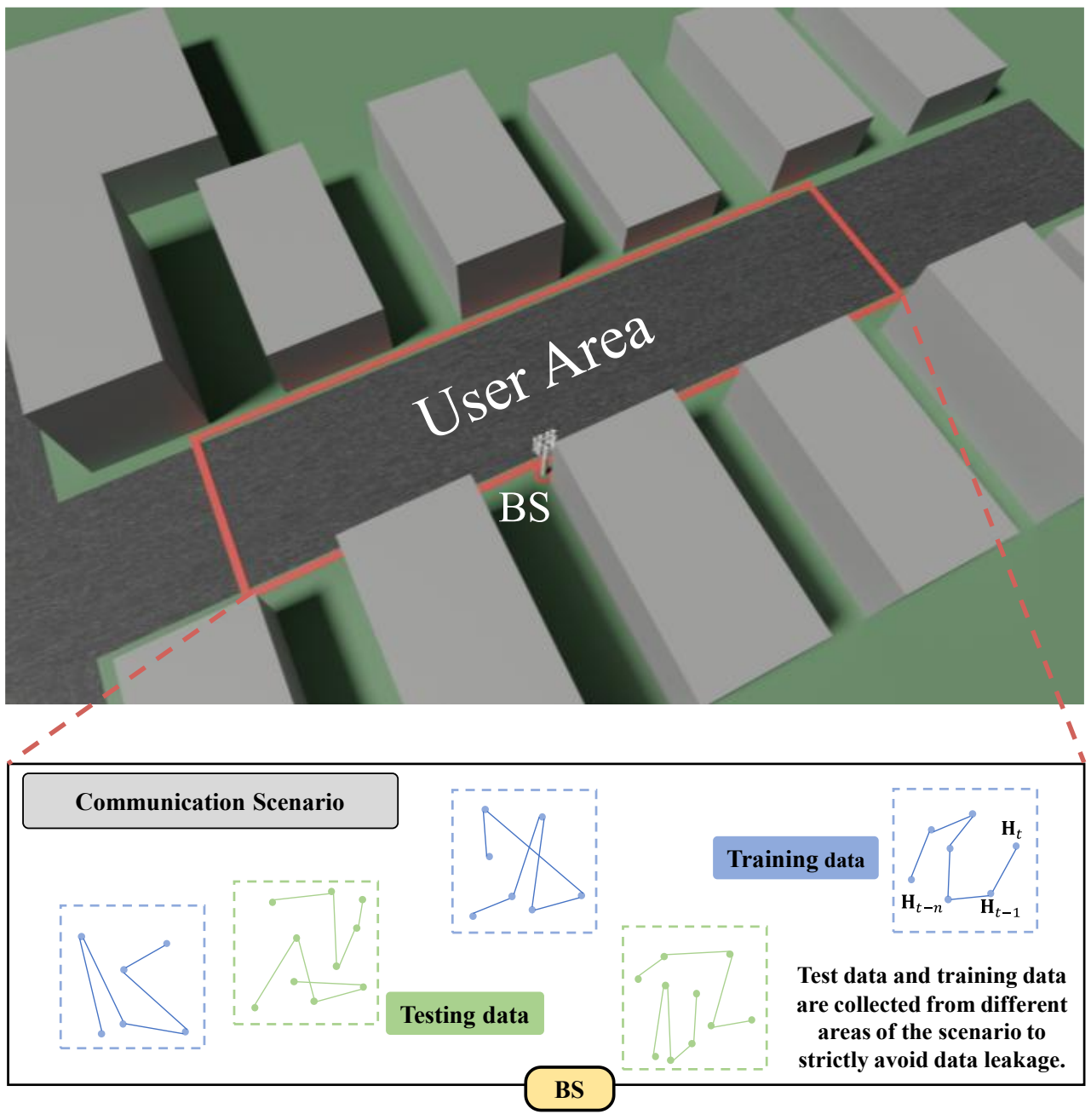

**Fig. 4**. Using the "O1" scenario in the DeepMIMO dataset [49] as experimental scenario and collecting training and testing datasets from it.

To prevent data leakage, training and testing data are collected from distinct sub-areas within the scenario, as illustrated in the bottom part of Fig. 4. As described in Section 3.3, we perform sequence shuffling on the channel samples from the training sub-areas to achieve training data augmentation. Simultaneously, we generate a fixed number of channel sequences from the testing sub-areas for model evaluation. The training and testing data include both quasi-static and high-mobility scenarios to simulate different user motion patterns. In each channel sequence, the last channel is designated as the present channel, while the previous ones are treated as past channels. The acquisition results of the present channel from all schemes will be compared with the ground truth for performance assessment. Table 2 lists the detailed simulation configurations of the DeepMIMO dataset.

**Table 2**

Default parameter settings for the DeepMIMO dataset.

| Parameters | Value |
| --- | --- |
| Frequency band | 3.5 GHz |
| Bandwidth | 40 MHz |
| BS index in "O1" scenario | BS-3 |
| Antenna array form | ULA |
| Number of antennas ($N_t$) | 32 |
| Number of subcarriers ($N_c$) | 32 |
| Number of paths ($P$) | 25 |
| User area | R501–R1400 |
| Number of training sub-areas | 2000 |
| Number of channels collected in each training sub-area | 32 |
| Number of testing sub-areas | 1000 |
| Number of testing sequences | 40 000 |
| Sequence length of each piece of testing data | 17 (16 past channels and 1 present channel) |

### *4.1.2 Performance indexes*

We use the normalized MSE (NMSE) and cosine correlation $\rho$ [27] between the acquired channel and true channel as the performance indexes, which are defined as follows:

$$\text{NMSE} = \mathbb{E}\left\{\frac{\left\|\mathbf{H}_{t_0} - \widehat{\mathbf{H}}_{t_0}\right\|_F^2}{\left\|\mathbf{H}_{t_0}\right\|_F^2}\right\} \tag{20}$$

and

$$\rho = \mathbb{E}\left\{\frac{1}{N_c}\sum_{n=1}^{N_c}\frac{\left|\hat{\mathbf{h}}_n^{\mathrm{H}}\mathbf{h}_n\right|}{\left\|\hat{\mathbf{h}}_n\right\|_2\left\|\mathbf{h}_n\right\|_2}\right\} \tag{21}$$

where $\mathbf{h}_n$ and $\hat{\mathbf{h}}_n$ are the original and acquired CSI of the $n$th subcarrier (i.e. the $n$th columns of the CSI matrices $\mathbf{H}_{t_0}$ and $\widehat{\mathbf{H}}_{t_0}$), respectively.

### *4.1.3 Benchmarks and training settings*

To evaluate the performance of the proposed Coupler architecture in the channel deduction task, we adopt two types of representative baseline models. One is the vanilla CDNets proposed in Ref. [42], which includes two specific implementations: the recurrence-based RCDNet and the attention-based ACDNet. The other is ConvLSTM [50], a classical image sequence model that replaces the linear layers within the LSTM with convolutional operations, processing channel sequences as dual-channel image sequences. For the three Coupler models proposed in this paper, the model depth is set to $K = 8$, with hidden dimensions $S_t = S_c = 128$. All models are trained for $10^6$ epochs and optimized using the adaptive moment estimation (Adam) optimizer [51]. For ACDNet and RCDNet, we follow the parameter settings in Ref. [42]. Additionally, we construct two enlarged CDNet variants—namely, ACDNet-Big and RCDNet-Big—in which the hidden width $S$ of the temporal learning module is increased from 512 to 1024. Table 3 lists the number of parameters and the FLOPs for all models (the case of the known present channel size is $3$ antennas $\times$ 3 subcarriers, and the number of past channels is 16). As a compact architecture tailored to the intrinsic structure of wireless channels, Coupler requires substantially fewer parameters than the vanilla CDNets and offers significant advantages over ConvLSTM in both computational complexity and parallel efficiency.

**Table 3**

Parameter counts and FLOPs of models.

| Model | Parameter counts | FLOPs |
| --- | --- | --- |
| ConvCoupler | 313.46 K | 294.68 M |
| ACoupler | 513.02 K | 521.75 M |
| AFCoupler | 513.02 K | 506.54 M |
| RCDNet | 6.53 M | 194.88 M |
| ACDNet | 11.79 M | 373.49 M |
| RCDNet-Big | 21.22 M | 660.63 M |
| ACDNet-Big | 42.24 M | 1.37 G |
| ConvLSTM | 42.37 K | 1.01 G |

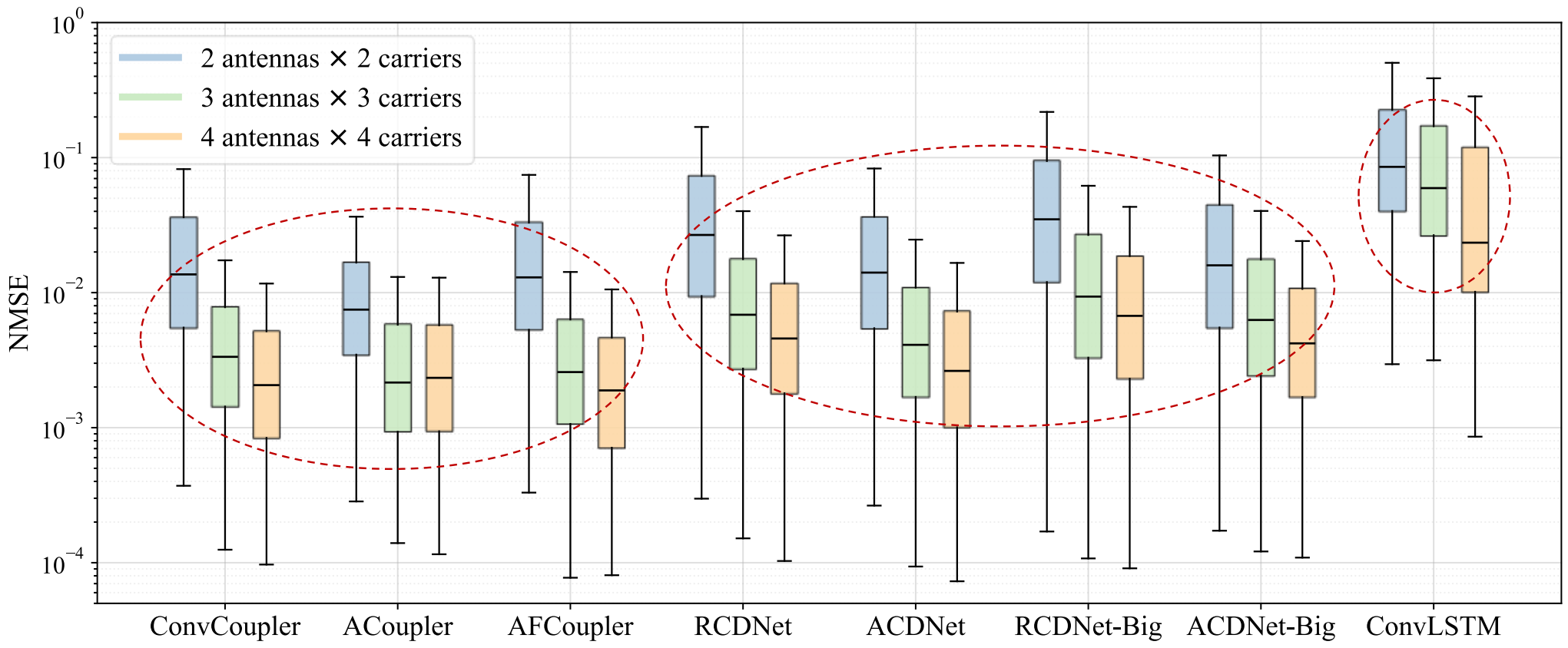

**Fig. 5.** NMSE of proposed Couplers and benchmarks under various estimated present channel sizes. The sizes of the estimated present partial channels are shown in the legend, and the size of the full channel is 32 antennas × 32 subcarriers. In each box, the central line indicates the median, the box edges represent the 25th and 75th percentiles, and the whiskers extend to the furthest data points within 1.5 times the interquartile range (IQR).

## 4.2 Performance evaluation

### *4.2.1 Accuracy of acquired CSI*

The accuracy of the acquired present CSI serves as a critical metric for evaluating the performance of the channel deduction models. Fig. 5 and Table 4 present the NMSE box plots and the average cosine similarity $\rho$, respectively, across different sizes of the known present channel. The full channel size is $32 \times 32$, and the past channel sequence length is set to 16. Compared with ConvLSTM, the Couplers and vanilla CDNets, which are specifically designed for channel deduction tasks, exhibit significant advantages in acquisition accuracy. This is attributed to the fact that wireless channels are complex tensors with distinct physical structures and lack the translation invariance characteristic of natural images. Consequently, existing generic image sequence processing models fail to handle them effectively. Furthermore, compared with the vanilla CDNets, the proposed Couplers achieve further improvements in accuracy. Across all sizes of known present channel, the proposed Coupler models outperform RCDNet and ACDNet in terms of both mean and median NMSE on the test set; they also achieve superior performance in cosine similarity $\rho$. Taking AFCoupler as an example, in the specific case of 3 antennas × 3 subcarriers, it achieves median and mean NMSE gains of 2.01 dB and 2.66 dB, respectively, over ACDNet. When benchmarked against RCDNet, the median and mean gains further increase to 4.24 dB and 4.51 dB. Therefore, compared with the existing CDNets, Couplers achieve higher accuracy with fewer parameters, indicating superior parameter efficiency. Even when CDNets are enlarged to exceed Couplers in computational cost, their performance shows no clear improvement, confirming that Couplers' advantage comes from the architectural design rather than increased computation.

**Table 4**

The cosine correlation $\rho$ of the proposed Couplers and benchmarks under various estimated present channel sizes.

| Estimated channel size through pilots | ConvCoupler | ACoupler | AFCoupler | ACDNet | ACDNet-Big | RCDNet | RCDNet-Big | ConvLSTM |
|---|---|---|---|---|---|---|---|---|
| 2 antennas × 2 carriers | 0.9844 | **0.9940** | 0.9878 | 0.9846 | 0.9809 | 0.9765 | 0.9717 | 0.9220 |
| 3 antennas × 3 carriers | 0.9968 | **0.9978** | 0.9975 | 0.9951 | 0.9896 | 0.9931 | 0.9905 | 0.9294 |
| 4 antennas × 4 carriers | 0.9975 | 0.9974 | **0.9978** | 0.9947 | 0.9922 | 0.9952 | 0.9932 | 0.9488 |

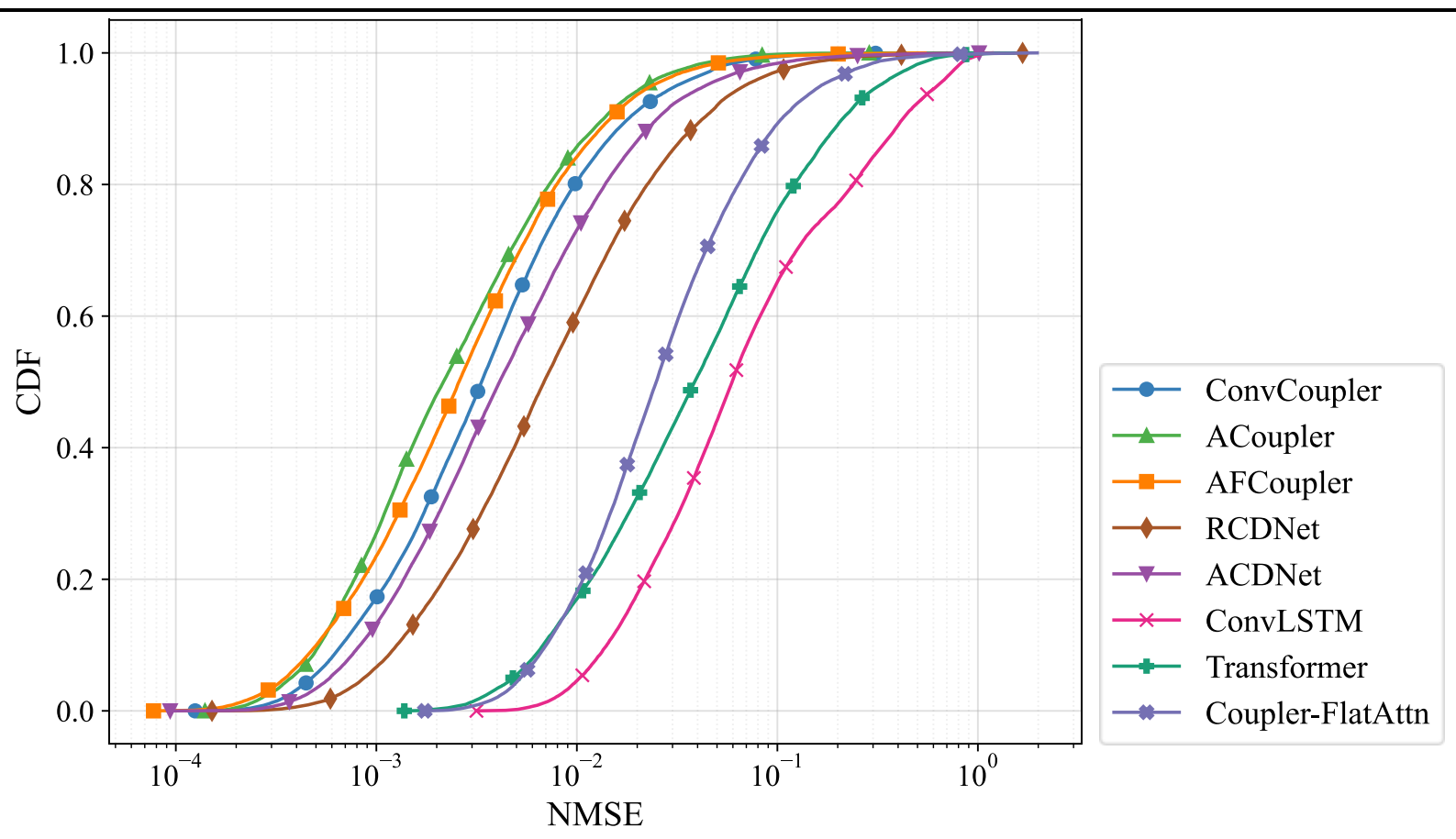

**Fig. 6.** Cumulative probability distribution of errors between the acquired channel and the true channel (NMSE).

Fig. 6 illustrates the cumulative distribution function (CDF) of the NMSE for different models on the test set. Here, the known present channel size is fixed at 3 antennas $\times$ 3 subcarriers, a configuration maintained in subsequent experiments. It can be observed that the NMSE for all three Coupler models is concentrated below 0.1, and the overall error distribution is superior to that of the vanilla CDNets. This indicates that the proposed models can provide high-quality channel acquisition services for users at nearly all locations within the scenario.

To further verify the advantages of Coupler's FDIL architecture and structure-preserving operators, we conduct two additional ablation studies. First, we adopt a standard Transformer model that flattens the 2D spatial-frequency CSI at each time slot, embeds it into a 1D feature vector through fully connected layers, and performs temporal modeling using a 12-layer Transformer with $d_{\text{model}} = 512$. As shown in Fig. 6, its NMSE distribution is less favorable than that of ACDNet, suggesting that temporal Transformer layers alone cannot sufficiently exploit the 2D spatial-frequency channel structure after flattening. This is because ACDNet sandwiches the temporal Transformer module between two spatial-frequency feature mixing modules, allowing the spatial-frequency structure to be preliminarily mapped before temporal interaction and further refined afterward. In contrast, Coupler goes one step further by interleaving the temporal, spatial, and frequency mixing layer by layer, thereby enabling more fine-grained full-domain feature fusion throughout the entire representation process.

Second, we construct Coupler-FlatAttn to validate the necessity of the IC-2D projection mechanism in ACoupler and AFCoupler. In this ablation variant, the IC-2D projections for generating Q/K/V in the temporal mixing layer are replaced by conventional 1D projections based on flattened features, and standard attention is used for temporal interaction. Compared with ACoupler and AFCoupler, this flattened-attention variant suffers from a degraded NMSE distribution while incurring substantially higher parameter count and computational complexity, indicating that directly flattening the 2D spatial-frequency CSI both disrupts the interlaced structure of the channel features and leads to redundant high-dimensional projections during temporal mixing. Therefore, IC-2D projection is not only beneficial for reducing parameter count and computational complexity but also essential for preserving the spatial-frequency channel structure and enabling efficient full-domain interleaved learning in Coupler.

### *4.2.2 Convergence during training*

Fig. 7 illustrates the average NMSE on the test set for all models during the training process. All models were sufficiently trained, showing a steady decrease in NMSE until convergence. Notably, the three Coupler models exhibited the fastest convergence rates and achieved the best NMSE performance upon convergence. Compared with the baseline models, the Couplers require a significantly shorter training time to attain satisfactory performance, which offers advantages for time-sensitive applications such as online learning and fine-tuning.

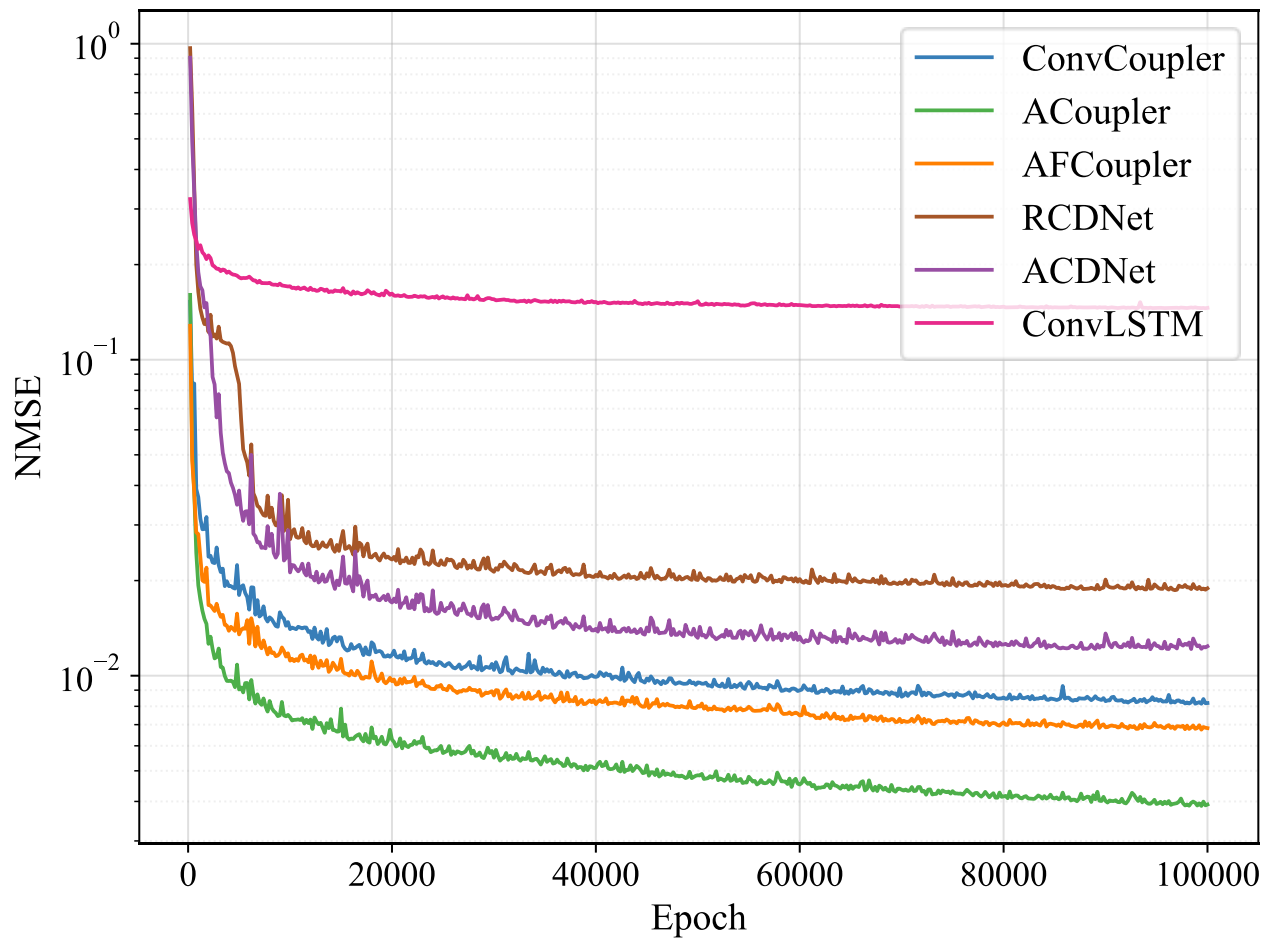

**Fig. 7**. NMSE convergence curves on the test set during training.

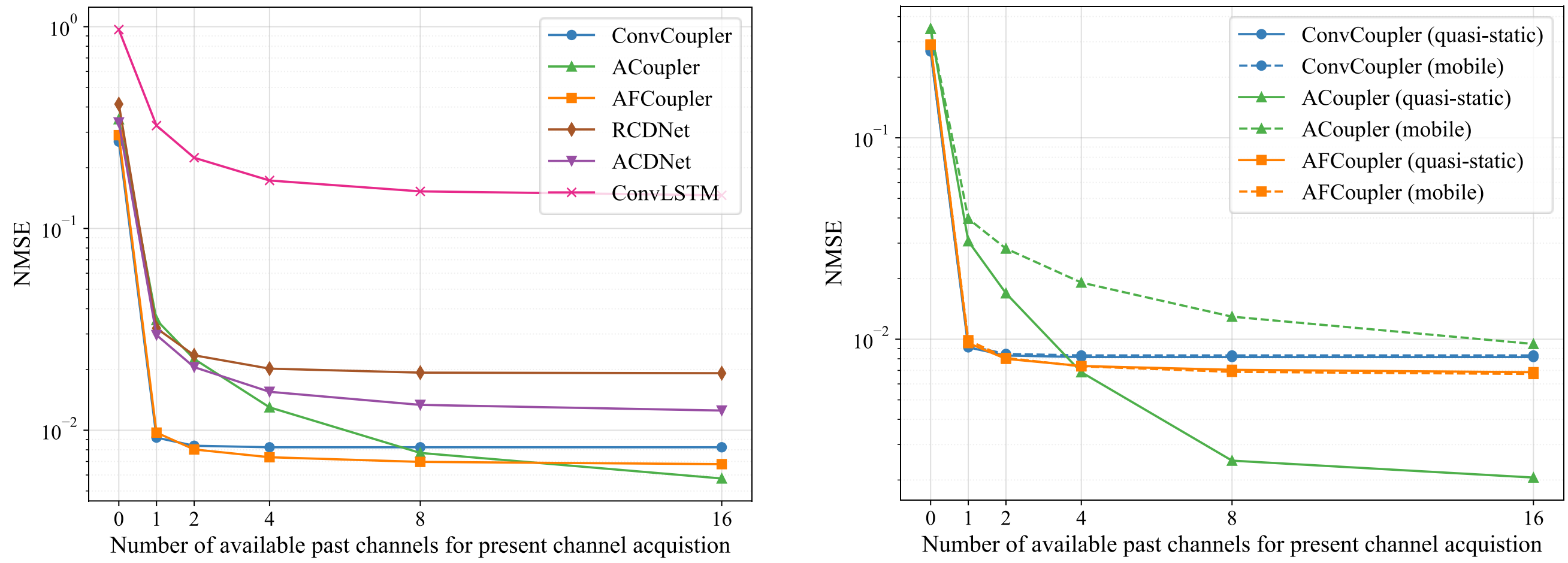

(a) (b)

**Fig. 8.** Impact of the number of available past channels on channel deduction accuracy. The size of the estimated partial channel through pilots is 3 antennas × 3 subcarriers. (a) NMSE vs. the number of available past channels for all models; (b) performance of Couplers under different user motion patterns.

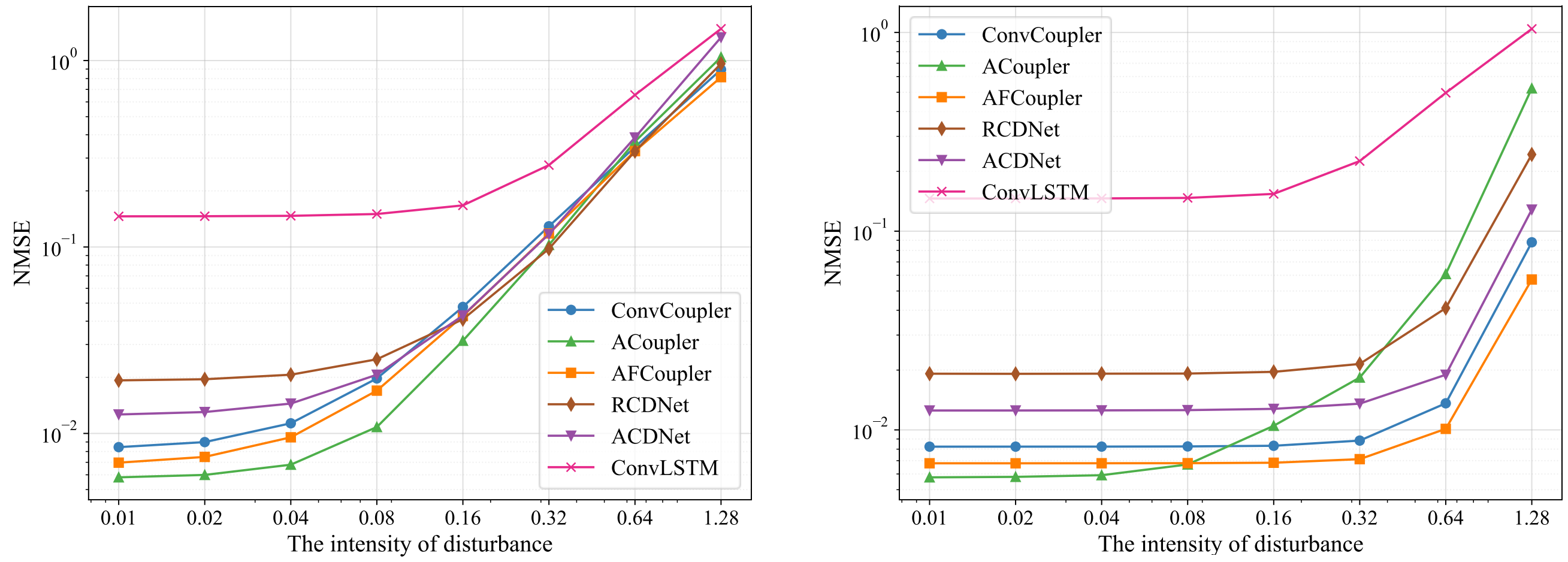

(a) (b)

**Fig. 9.** Robustness evaluation of different models. The size of the estimated partial channel through pilots is 3 antennas × 3 subcarriers. (a) NMSE when different disturbance intensities $\sigma$ are added to a past and known partial present channel; (b) NMSE under only the addition of different disturbance intensities $\sigma$ to a past channel.

#### *4.2.3 Acquired CSI accuracy vs. number of past channels*

The advantage of channel deduction over channel estimation primarily lies in the extraction of additional information from past channels to facilitate the acquisition of the present CSI. Fig. 8(a) illustrates the impact of the number of past channels on the accuracy of the acquired CSI. Within the channel deduction framework, the acquisition accuracy of all models improves as the number of past channels increases. Notably, ConvCoupler and AFCoupler can achieve high acquisition accuracy even with a limited number of past channels. In the case of a single past channel, they exhibit a performance gain of over 4.5 dB compared with the vanilla CDNets. Specifically, the performance of ConvCoupler saturates when the number of past channels $N > 4$, which is attributed to the limited receptive field of the 1D convolutional kernels. In contrast, ACoupler excels at leveraging long sequences of past channels to achieve performance gains. This advantage is derived from the global receptive field of the attention mechanism, which enables ACoupler to capture fine-grained channel correlations across long temporal sequences.

To further investigate the features of the three Coupler derivatives under different user motion patterns, Fig. 8(b) illustrates the performance of the Couplers in both quasi-static and mobile cases. It can be observed that ConvCoupler and AFCoupler exhibit comparable performance in these two cases, whereas ACoupler excels in the quasi-static case but performs slightly worse under high-mobility conditions. This is attributed to the fact that ACoupler is capable of directly extracting the information most correlated with the present channel from past channel sequences, making it more suitable for quasi-static scenarios. Conversely, ConvCoupler and AFCoupler employ convolution operations and sequence feature fusion mechanisms, respectively, which enable them to learn global channel structure from historical data, making them robust to mobility pattern variations.

#### *4.2.4 Robustness against channel disturbances*

In practical scenarios, influenced by noise and interference, the past channels and the present partial channel fed into the model are imperfect. This sub-section evaluates the robustness of the channel deduction models against input channel disturbances, considering two distinct cases. The first case assumes that all inputs $\{\mathbf{H}_{t_0-N}, \dots, \mathbf{H}_{t_0-1}, \mathbf{H}_{t_0}[\Omega]\}$ are lossy, aiming to test the model's capability to withstand input noise. The second case considers a scenario in which the past channels $\{\mathbf{H}_{t_0-N}, \dots, \mathbf{H}_{t_0-1}\}$ are imperfect while the known present subchannel $\mathbf{H}_{t_0}[\Omega]$ is perfect, aiming to evaluate the model's self-calibration capability to leverage current new observations to mitigate the impact of historical errors. The channel disturbance is introduced as follows [39]: $\mathbf{H}_{\text{dis}} = \mathbf{H} \odot \mathbf{D}$, where $\mathbf{H}_{\text{dis}}$ is the disturbed CSI matrix, $\mathbf{H}$ is the ideal CSI matrix, $\mathbf{D}$ is the disturbance matrix, and each element of $\mathbf{D}$ is an independent and identically distributed Gaussian random variable obeying $\mathcal{N}(1, \sigma^2)$. $\sigma$ simulates the degree of deviation of the lossy channel from the ideal channel.

Fig. 9(a) illustrates the relationship between model performance and the disturbance intensity $\sigma$ when both past and current channels are imperfect. In the low-disturbance regime ($\sigma < 0.08$), the three Coupler models exhibit robust performance; under larger disturbances, however, the performance of all models degrades rapidly. Fig. 9(b) analyzes the model performance when disturbances are applied exclusively to the past channels. Compared with Fig. 9(a), the robustness of all models against high-intensity disturbances is significantly improved in this case. This result underscores the importance of an accurate current partial channel for channel acquisition, which provides critical small-scale fast-fading information. Among the proposed models, AFCoupler and ConvCoupler demonstrate superior robustness, while ACoupler exhibits relatively higher sensitivity. This result is attributed to the ACoupler's reliance on fine-grained feature correlations between past channels and present subchannels for channel deduction: While it achieves a higher upper bound on channel acquisition accuracy in noise-free scenarios, it is also more susceptible to channel disturbances. These experimental results validate the outstanding data efficiency of the Coupler's FDIL architecture in wireless channel representation tasks.

#### *4.2.5 Effect of training data volume*

Fig. 10 illustrates the test performance of each model when trained with varying numbers of sub-areas. It can be observed that the proposed Coupler models exhibit a more substantial performance advantage in the situation with limited training data. When the number of training sub-areas is reduced to 1000, AFCoupler achieves a mean NMSE performance gain of over 4.75 dB and a median NMSE gain of 6.7 dB compared with the vanilla CDNets. This result is attributed to the FDIL architecture of the Couplers, which introduces a stronger inductive bias and significantly reduces the number of model parameters, thereby enabling superior generalization performance under low-data conditions. As the volume of training data increases, the performance of all models consistently improves. In the sufficient data regime (3000 training sub-areas), ACoupler and AFCoupler still maintain a certain performance edge, while ConvCoupler exhibits slightly inferior performance compared with ACDNet due to its overly simplified time-domain mixing mechanism.

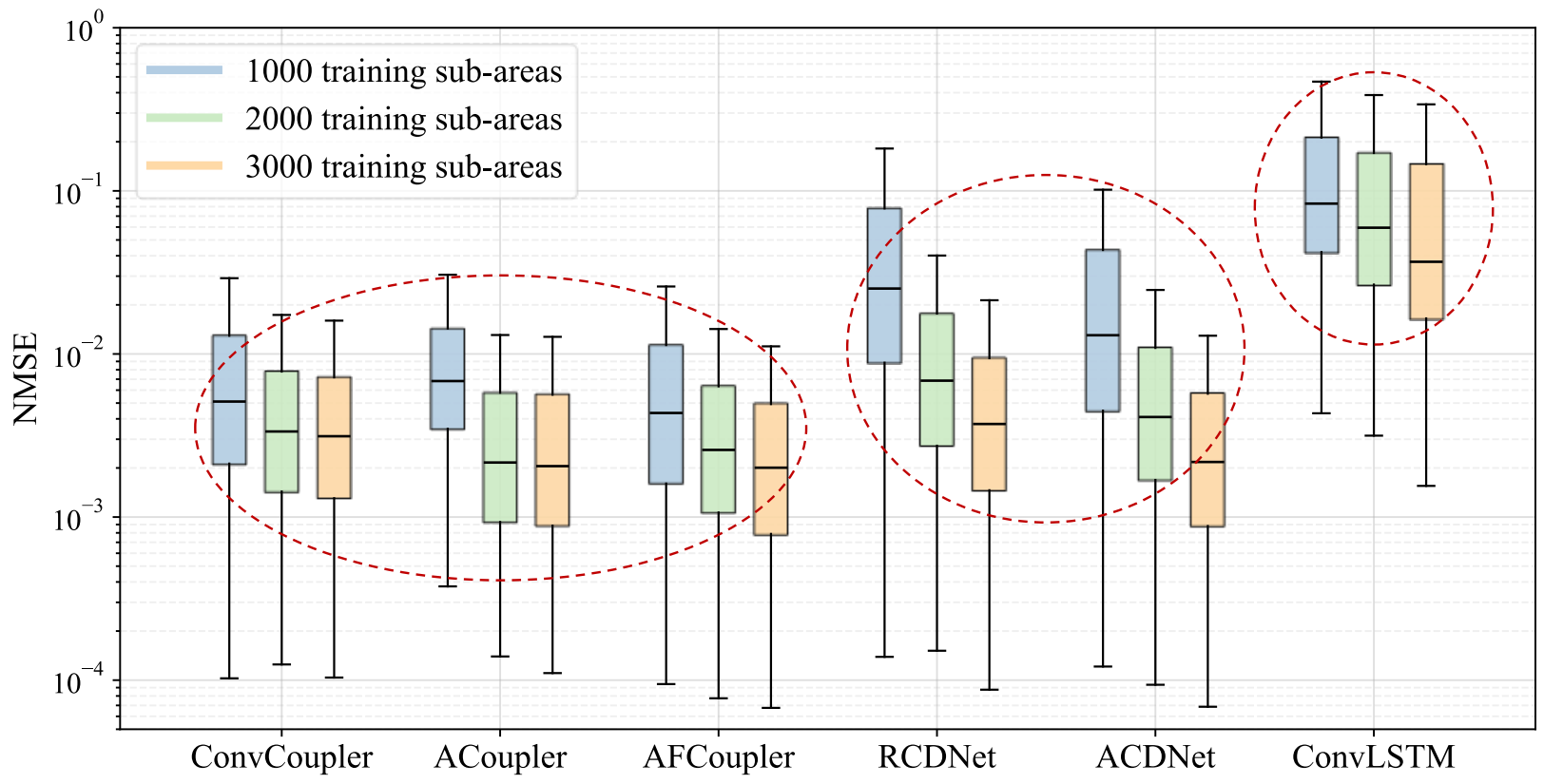

**Fig. 12.** NMSE of proposed Couplers and benchmarks under various numbers of training sub-areas.

#### *4.2.6 Continuous channel acquisition with Couplers*

Following the autoregressive deployment introduced in Section 3.3, we utilize the trained Coupler models to provide continuous channel acquisition services for a mobile user, aiming to evaluate the models' effectiveness in the presence of error propagation. Fig. 11(a) depicts the mobile user and its corresponding trajectory. The user equipment acquires complete CSI using high-density pilots during the initial $N$ time slots ($N = 8$), with this trajectory segment indicated by the green line. For the subsequent 2024 time slots, the user continues moving (including intermittent stationary periods), which is represented by the orange line. Throughout this process, the Coupler models continuously provide channel deduction services based on the known CSI of the preceding $n$ time slots and a partial estimate (3 antennas $\times$ 3 subcarriers) of the current channel.

Fig. 11(b–d) illustrates the NMSE of the three Coupler models across these 2024 time slots. It can be observed that all three models remain functional along the entire trajectory. Compared with the ideal case without error propagation, ACoupler exhibits substantial performance degradation in autoregression. Even in the ideal case, the NMSE of the channels obtained by ACoupler shows noticeable fluctuations. As discussed in Section 0, ACoupler relies on fine-grained temporal similarities to select highly relevant historical channel features. While this mechanism is effective in quasi-static or error-free cases, it is more sensitive to small-scale sample fluctuations and accumulated errors. Therefore, in the autoregressive process, the reused deduced channels may disturb the attention weights and further amplify error propagation. In contrast, the autoregressive channel acquisition accuracy of AFCoupler and ConvCoupler closely approaches that of the ideal error-free baseline. This is because AFCoupler and ConvCoupler focus more on long-term temporal structures and common features rather than point-wise extracting features from individual historical slots, making them less susceptible to local channel perturbations and historical estimation errors.

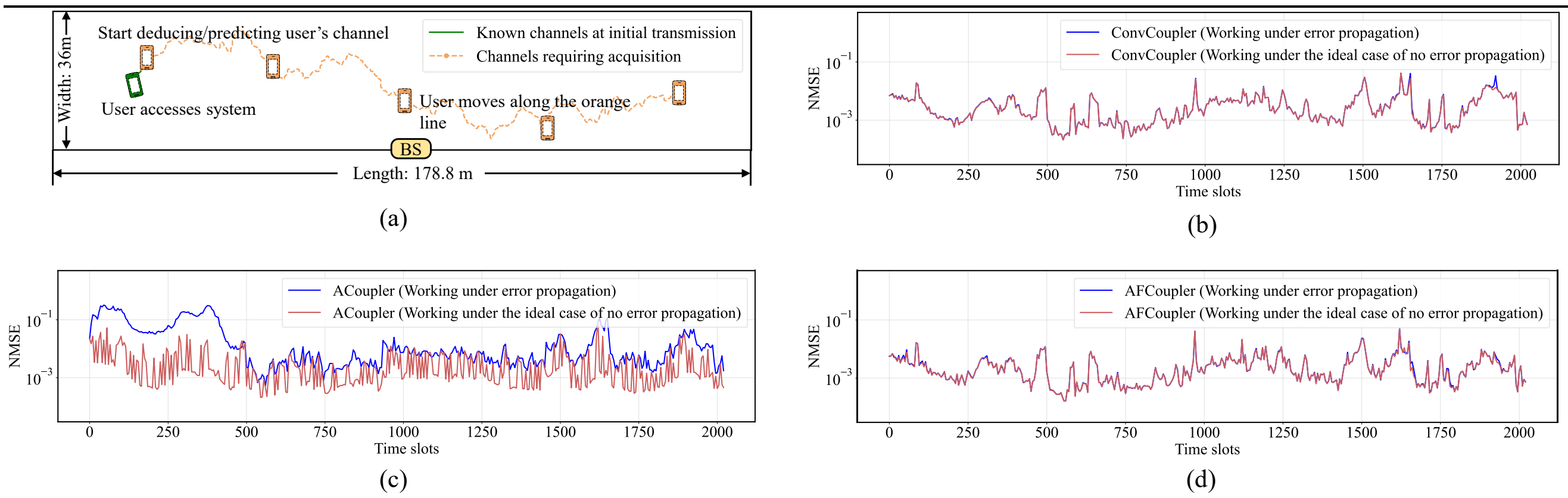

**Fig. 13.** Continuous channel acquisition for a mobile user using Couplers in the autoregressive scenario and the ideal case without error propagation. (a) Illustration of a mobile user's trajectory and the channel deduction process in the "O1" scenario; (b) NMSE along the mobile-user trajectory for ConvCoupler; (c) NMSE along the mobile-user trajectory for ACoupler; (D) NMSE along the mobile-user trajectory for AFCoupler.

### 4.3 Experiments on real-world data

To further validate the effectiveness of the channel deduction scheme driven by the proposed Couplers in practical wireless systems, this section presents experimental results on real-world data. Unlike ray-tracing-based simulation data with ideal samples or labels, measured CSI is obtained through pilot-based channel estimation and is inevitably affected by various system non-idealities and estimation errors. In this part, we adopt the DIstributed CHAnnel Sounder by University of Stuttgart (DICHASUS) [52], an open-source large-scale measured MIMO CSI dataset, for experiments. DICHASUS includes CSI measurements across various scenarios, and the CSI data are labeled with positioning coordinates and the SNR of each antenna channel. Here, we select the DICHASUS-015x dataset, which represents an indoor mobile communication scenario featuring scatterers such as metal plates, whiteboards, and walls, as shown in Fig. 12(a). This dataset consists of channel sampling points collected along six robot movement trajectories, labeled 0152, 0153, 0154, 0155, 0157, and 0158. Fig. 12(b) visualizes the sampling points in this scenario and the average SNR at corresponding locations. We sample 72 000 points from five trajectories (DICHASUS-0152 to 0157) to construct the training and testing sets. For each point, the 32 nearest points are selected to construct the channel sequence, maintaining the identical sequence construction method as before. To learn from the more challenging real-world data and mitigate the impact of data noise and non-ideal system effects, we expand the hidden-layer widths $S_t$ and $S_c$ of the CMLPs in the Coupler models to 256 and introduce dropout with a probability of 0.1 for all models in order to suppress overfitting to noisy labels and improve generalization on measured data.

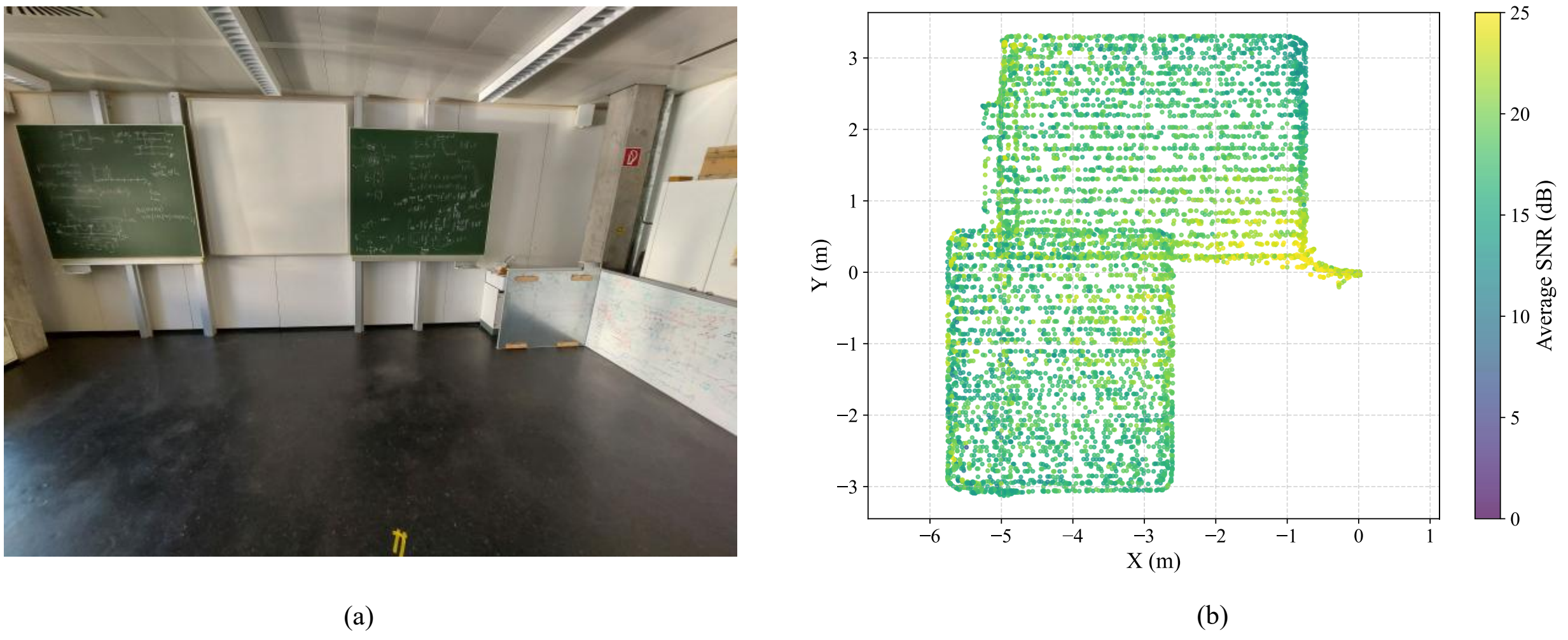

**Fig. 14.** Illustration of the DICHASUS-015x dataset [52]. (a) Scene photograph; (b) channel measurement locations and SNR.

Fig. 13 evaluates the NMSE performance across different schemes with respect to the size of the partial estimated present channel. Here, the size of the complete channel remains at 32 antennas × 32 subcarriers. Considering the more severe non-ideal characteristics and complex dynamics inherent in real-world measurements, we appropriately increase the real-time pilot sizes to provide sufficient calibration information. We introduce the classical linear minimum mean squared error (LMMSE) channel estimation [53] algorithm for comparison. It can be observed that, on the real-world data, the channel deduction models exhibit a substantial performance advantage over LMMSE. This is because LMMSE is essentially an SNR-adaptive linear estimator that mainly relies on current pilot observations and stable second-order channel statistics, making it less capable of capturing complex nonlinear channel variations in practical mobile scenarios. In contrast, through neural networks, the channel deduction models jointly leverage historical channel samples and current sparse pilots to learn specific state evolution and more effectively exploit the nonlinear feature correlations of channels across spatial, temporal, and frequency domains, thereby achieving significant performance improvements.

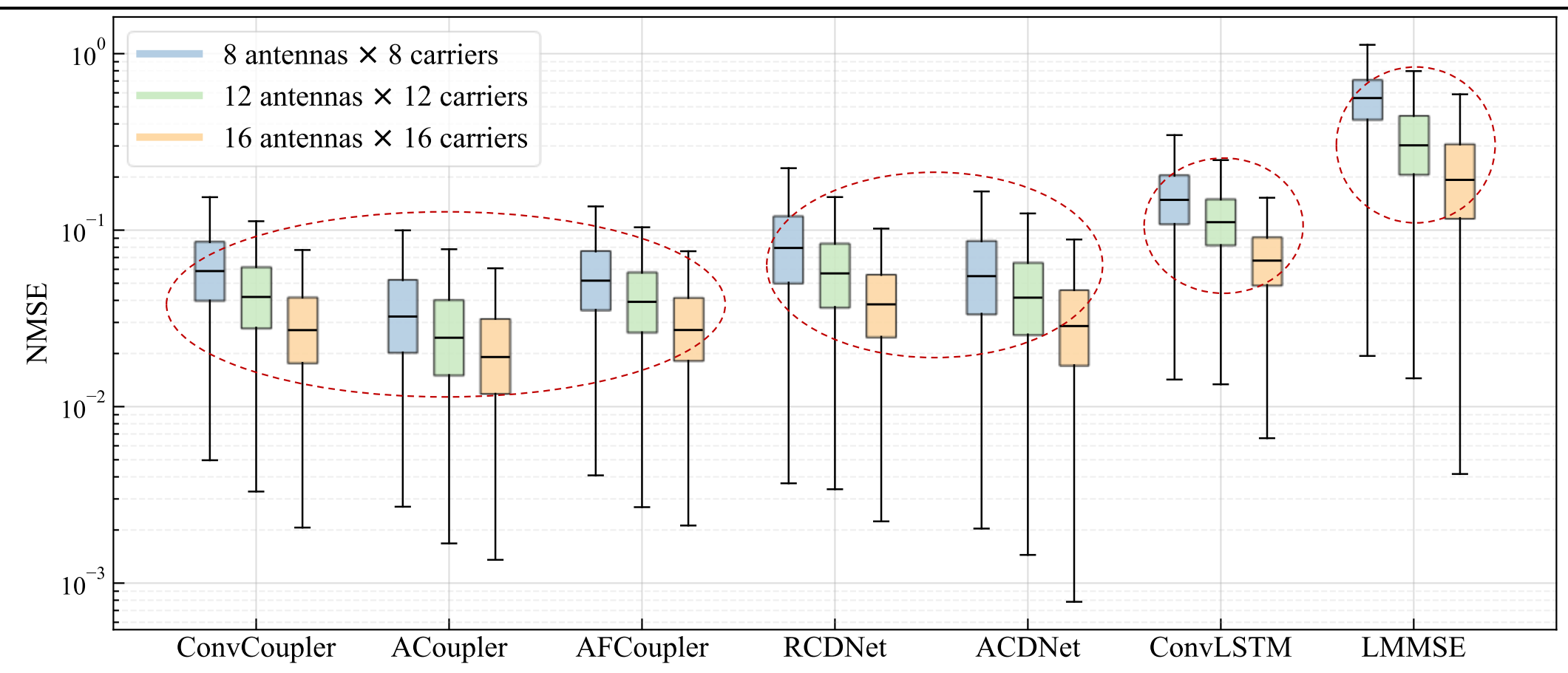

**Fig. 15.** Channel acquisition performance of different models under varying sizes of the estimated present channel. The models are tested using real-world channel measurements from trajectories DICHASUS-0152 to -0157.

Among the channel deduction models, ConvLSTM exhibits poor performance due to a mismatch between its network architecture and the characteristics of wireless channels. Conversely, both Couplers and the vanilla CDNets achieve favorable channel acquisition results. With more sophisticated structural designs, Couplers can achieve comparable or even superior channel acquisition performance while using fewer model parameters. Overall, the relative performance of the models on the real-world data aligns closely with the aforementioned simulation results. Essentially, although the rationale analysis of Coupler in Section 3.1 relies on classical mathematical models, the fundamental multi-domain coupling properties remain consistent in measured data. Consequently, Coupler leverages its inherent learning capacity to bridge the gap between theoretical modeling and physical reality, maintaining high efficiency across both synthetic benchmarks and practical evaluations.

To further demonstrate the practical viability of Coupler-based channel deduction in real-world systems, we employ AFCoupler as a representative example to perform autoregressive channel acquisition on a new trajectory, DICHASUS-0158, within this scenario. Fig. 14(a) illustrates the NMSE of the channel acquired by AFCoupler on this trajectory with 16 past channels. As observed, the NMSE of the channels acquired by the model remains below 0.1 at most locations, indicating good channel acquisition quality. Fig. 14(b) visualizes the NMSE at each time slot when AFCoupler performs channel deduction on this trajectory with varying numbers of past channels, where ACDNet, RCDNet, and the LMMSE algorithm are introduced as baselines. The experimental results show that both the CDNets and AFCoupler significantly outperform the conventional LMMSE channel estimation method in terms of channel acquisition performance, demonstrating the substantial advantages of deep-learning-based channel deduction schemes in practical wireless systems. Moreover, when the same number of past channels is used, AFCoupler consistently outperforms ACDNet and RCDNet throughout the entire autoregressive channel acquisition process. Notably, even with only $N = 4$ past channels, AFCoupler still surpasses ACDNet and RCDNet using $N = 16$ past channels. As more past channel information is introduced, the channel acquisition accuracy of AFCoupler gradually improves, ultimately approaching the performance upper bound under the ideal case without error propagation. These results highlight the importance of exploiting past channel information for current channel acquisition in practical systems, while validating the effectiveness of the AFCoupler-based channel deduction mechanism on real-world data.

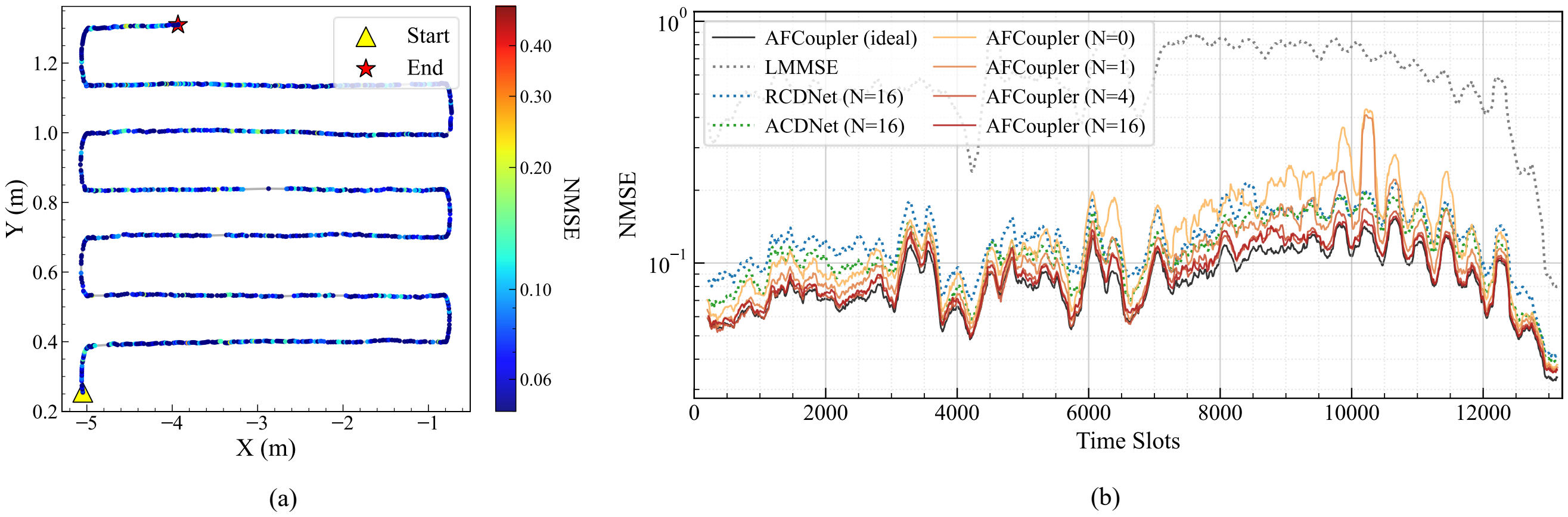

**Fig. 16.** Autoregressive channel acquisition performance of AFCoupler on a new trajectory, DICHASUS-0158 [52]. (a) NMSE of the autoregressively acquired channels along the trajectory with 16 past channels; (b) model performance in autoregressive channel deduction with different numbers of past channels, compared with the ideal case without error propagation ($N = 16$) and baseline models. A moving average filter with window size 200 is applied for smoothing.

## 5. Conclusion

This paper presented Coupler, a wireless-native backbone architecture for channel representation learning, from which is derived a diverse family of learning schemes including ConvCoupler, ACoupler, and AFCoupler. The unique full-domain interleaved learning mechanism and well-matched operators ensure its lightweight nature while enabling the highly effective extraction of complex temporal-spatial-frequency channel characteristics. Coupler's excellent structural usability and adaptability to continuously increasing learnable physical dimensions were demonstrated in detail through applications in channel deduction, a fundamental representation learning task. Comprehensive evaluations involving simulations, field measurements, and multidimensional metrics validated the effectiveness and superiority of the proposed architecture, complemented by an in-depth characteristic analysis of its various derivatives.

It is noteworthy that Coupler achieves superior performance not by further stitching emerging advanced networks but through a streamlined architecture. By simply stacking identical modules, it efficiently captures complex wireless knowledge using remarkably few neurons. This characteristic of deriving effectiveness from simplicity stems from Coupler's profound alignment with physical principles, further epitomizing the AI-native 6G vision: *Intelligence should simplify rather than complicate wireless systems*. Furthermore, the excellent scalability afforded by its simplicity positions Coupler as a promising learning architecture for emerging wireless foundation models. We hope that this backbone architecture, along with the wireless-native AI design philosophy it embodies, can inspire and support the development of more advanced applications in next-generation wireless networks.

## Acknowledgments

The authors acknowledge support in part from the National Natural Science Foundation of China under Grants 624B2129 and 62394292, and in part from the Fundamental Research Funds for the Central Universities under Grant 226-2024-00069.

## Compliance with ethics guidelines

Zirui Chen, Ziqing Xing, Zhaoyang Zhang, Hongning Ruan, Yuzhi Yang, Zhaohui Yang, Chongwen Huang, and Mérouane Debbah declare that they have no conflict of interest or financial conflicts to disclose.

## CRediT authorship contribution statement

**Zirui Chen:** Writing – review & editing, Writing – original draft, Methodology, Formal analysis, Investigation, Validation, Conceptualization, Funding acquisition. **Ziqing Xing:** Writing – review & editing, Writing – original draft, Formal analysis, Visualization, Data curation, Software, Investigation, Validation. **Zhaoyang Zhang:** Writing – review & editing, Supervision, Resources, Conceptualization, Funding acquisition, Project administration. **Hongning Ruan:** Writing – review & editing, Investigation, Data curation, Software. **Yuzhi Yang:** Writing – review & editing, Formal analysis. **Zhaohui Yang:** Writing – review & editing, Supervision. **Chongwen Huang:** Writing – review & editing, Supervision. **Mérouane Debbah:** Writing – review & editing, Supervision.

## Appendix A. Supplementary code

All implementations are available at https://github.com/XIronMan0220/Coupler-Channel-Deduction.